\documentclass{article}%
\usepackage{amsmath}
\usepackage{amsfonts}
\usepackage{amssymb}
\usepackage{graphicx}%
\providecommand{\U}[1]{\protect\rule{.1in}{.1in}}
\begin{document}

\title{Localization and the interface between quantum mechanics, quantum field theory
and quantum gravity\\{\small dedicated to the memory of Rob Clifton}\\{\small submitted to "Studies in History and Philosophy of Physics"}}
\author{Bert Schroer\\CBPF, Rua Dr. Xavier Sigaud 150 \\22290-180 Rio de Janeiro, Brazil\\and Institut fuer Theoretische Physik der FU Berlin, Germany}
\maketitle
\tableofcontents

\begin{abstract}
We show that there are significant conceptual differences between QM and QFT
which make it difficult to view the latter as just a relativistic extension of
the principles of QM. At the root of this is a fundamental distiction between
Born-localization in QM (which in the relativistic context changes its name to
Newton-Wigner localization) and \textit{modular localization} which is the
localization underlying QFT, after one liberates it from its standard
presentation in terms of field coordinates. The first comes with a probability
notion and projection operators, whereas the latter describes causal
propagation in QFT and leads to thermal aspects of locally reduced finite
energy states. The Born-Newton-Wigner localization in QFT is only applicable
asymptotically and the covariant correlation between asymptotic in and out
localization projectors is the basis of the existence of an invariant
scattering matrix.

Taking these significant differences serious has not only repercussions for
the philosophy of science, but also leads to a new structural properties as a
consequence of vacuum polarization: the area law for \textit{localization
entropy} near the the causal localization horizon and a more realistic cutoff
independent setting for the cosmological vacuum energy density which is
compatible with local covariance. The article presents some observations about
the interface between QFT in CST and QG.

\end{abstract}

\section{Introductory remarks}

Ever since QM was discovered, the conceptual differences between classical
theory and quantum mechanics (QM) have been the subject of fundamental
investigations with profound physical and philosophical consequences. But the
conceptual relation between quantum field theory (QFT) and QM, which is at
least as challenging and rich of surprises, has not received the same amount
of attention and scrutiny, and often the subsuming of QFT under "relativistic
QM" nourished prejudices and prevented a critical foundational debate. Apart
from some admirable work on the significant changes which the theory of
measurements must undergo in order to be consistent with the structure of QFT
\cite{Rob} and some deep mathematical related work related to it
\cite{Werner}, the knowledge on this subject has remained in the mind of a few
individuals working on the foundations of QFT.

Often results of this kind which involve advanced knowledge of QFT do not
attract much attention even when they have bearings on the foundations of QT
as e.g. the issue of \textit{Bell states} in \textit{local quantum physics}
(LQP\footnote{We use this terminology instead of QFT if we want to direct the
reader's attention away from the textbook Lagrangian quantization towards the
underlying principles \cite{Haag}. QFT (the content of QFT textbooks) and LQP
deal with the same physical principles but LQP is less comitted to a
particular formalism (Lagrangian quantization, functional integrals) and
rather procures always the most adaequate mathematical concepts for their
implementation. It includes of course all the results of the standard
perturbative Lagrangian quantization but presents them in a conceptually and
mathematically more satisfactory way. Most of the subjects in this article are
outside of textbook QFT.}) \cite{Su} or the important relations between causal
disjointness with the existence of uncorrelated states as well as the issue to
what extend causal independence is a consequence of statistical independence
\cite{Bu-Su}. The reason is not so much a lack of interest but rather that QFT
is often thought to be just a kind of relativistic quantum mechanics. This may
explain why there has been a tremendous effort on the side of quantum
mechanical foundations and very little investment on the side of QFT. Indeed
there is an amazing lack of balance between the very detailed and
sophisticated literature about interpretational aspects of QM and its relation
with information theory, aiming sometimes at some very fine, if not to say
academic/metaphoric points (e.g. the multiworld interpretation), and the
almost complete lack of profound interpretive activities about our most
fundamental theory of matter. Although the name QT usually appears in the
title of foundational papers, this mostly hides the fact that they deal
exclusively with concepts from QM leaving out QFT.

If on the other hand some foundational motivated quantum theorist become aware
of the deep conceptual differences between particles and fields they tend to
look at them as antagonistic and create a battleground; the fact that they are
fully compatible where for physical reasons they must agree, namely in the
asymptotic region of scattering theory, is usually overlooked. 

The aim of this essay to show that at the root of these differences there are
two localization concepts: the quantum mechanical Born-Newton-Wigner
localization and the modular localization of LQP. The B-N-W localization is
not Poincar\'{e} covariant but attains this property in a certain asymptotic
limit namely the one which is needed in scattering theory. Modular
localization on the other hand is causal at all distances but lacks projectors
on subspaces, the linear spaces of localized states are usually dense in the
Hilbert space of all states. One of the aims of this article is to collect
some facts which show that besides sharing the notion of Hilbert space,
operators and states as well as $\hslash,$ QM and QFT are conceptually worlds
apart and yet they harmonize perfectly in the asymptotic region of scattering theory.

In this connection one is reminded that some spectacular misunderstanding of
conceptual properties in passing from QM to LQP led to incorrect results about
alleged violations of the velocity of light remaining a limiting velocity in
the quantum setting (the famous Fermi Gedankenexperiment). As a result of a
publication in Phys. Lett. \cite{Heger} and a simultaneous article in
\textit{Nature} on the prospects of time machines, this created quite a stir
at the time and led to a counter article \cite{Bu-Yn}. Since the LQP
presentation of the Fermi Gedankenexperiment has been a strong motivations for
non-experts to engage with its conceptual setting, and hence has some
pedagogical merits in the present context, it is natural that it will also
obtain some space in this article. Although these kinds of sophisticated
misunderstanding continue to appear occasionally in papers, only the mentioned
episode made it into the world press (as a result of the impact of PLR
articles on popular scientific journals as Nature and on the international press.)

It is not our intention to present a new axiomatic setting (for an older
presentation see \cite{Haag}). Such a goal would be too ambitious in view of
the fact that we are confronting a theory where, in contradistinction to QM,
no conceptual closure is yet in sight. Although there has been some remarkable
nonperturbative progress concerning constructive control (i.e. solving the
existence problem) of models, the main knowledge about models of QFT is still
limited to numerically successful but nevertheless diverging perturbative series.

Here the more modest aim is to collect some either unknown or little known
facts which could present some food for thoughts about a more inclusive
measurement theory, including all of quantum theory (QT) end not just QM. On
the other hand one would like to improve the understanding about the interface
between QFT in CST (curved spacetime) and the still elusive QG. This can only
be achieved by going somewhat beyond the presently fashionable "shut up and
calculate" attitude. But if one has to enter speculative excursions one would
like to do this from a solid conceptual-mathematical platform, so in case the
trip into the unknown ends in nowhere, there is a return and/or a chance to
modify the direction\footnote{In my lifetime I have seen 3 TOEs (theories of
everything) fail and a fourth is already in an intensive care unit. It is
common to all these attempts at a TOE that none of them were started from a
solid conceptual platform and only a few of their fans succeeded to return to
secure areas of particle physics.}.

Since both expressions QFT and LQP are used do denote the same theory, let me
emphasize again that there is no difference in content between; LQP is used
instead of QFT whenever the conceptual level of the presentations gets beyond
that which the reader is able to find in standard textbooks of QFT. There is
of course one recommendable exception, namely Rudolf Haag's book "Local
Quantum physics" \cite{Haag}; but in a fast developing area of particle
physics two decades (referring to the time it was written) are a long time. \ 

The paper consists of two main parts, the first is entirely dedicated to the
exposition of the differences between (relativistic\footnote{In order to show
that making QM relativistic does not remove the fundamental differences with
QFT the next section will be on the relativistic setting of "direct particle
interactions".}) QM and LQP, whereas the \ second deals with thermal
consequences of vacuum polarization caused by causal localization and some
consequences for QFT in curved spacetime (CST). A QG theory does not yet
exist, but a profound understanding of those foundational aspects are expected
to be important to get there.

The first part starts with a subsection on \textit{direct particle
interactions }(DPI), a framework which incorporates all those \ properties of
a relativistic theory which one is able to formulate solely in terms of
relativistic particles; some of them already appearing in the S-matrix work of
E.C.G. St\"{u}ckelberg. However the enforcement of the cluster factorization
property (the spatial aspect of macro-causality) in DPI requires more involved
arguments. It is not automatic as in nonrelativistic QM where it follows from
the additivity of interaction term. As a result DPI does not allow a second
quantization presentation, even though it is a perfect legitimate
multiparticle theory in which n-particles are linked to n+1 particles by
cluster factorization. Within the particle physics community there seems to be
a lack of awareness about its existence which may be due to the fact that its
protagonists are theoretical nuclear physicists who wanted to construct a
relativistic particle theory for an intermediate energy range for which
relativistic invariance is already important but only a few particles can be
created. Most particle physicists tend to believe that a relativistic particle
theory, consistent with macro-causality and a Poincar\'{e}-invariant S-matrix,
must be equivalent to QFT\footnote{The related folklore one finds in the
literature amounts to the dictum: relativistic quantum theory of particles +
cluster factorization property = QFT. Apparently this conjecture goes back to
S. Weinberg.}, therefore it may be helpful to show that this is not correct.

Since the ideas which go into its construction are important for appreciating
the conceptual differences of relativistic QM to QFT, we will at least sketch
some of the arguments showing that DPI theories fulfill all the physical
requirements which one is able to formulate solely in terms relativistic
particles without recourse to fields, as Poincar\'{e} covariance, unitary and
macro-causality of the resulting S-matrix (which includes cluster
factorization). In contradistinction to nonrelativistic mechanics for which
clustering follows trivially from the additivity of pair-(or higher-) particle
potentials, and also in contradistinction to QFT where the clustering is a
rather straightforward consequence of locality and the energy positivity, the
implementation in the relativistic DPI setting is much more subtle on this
point (and this is related to the lack of a second quantization reformulation
of multi-particle interactions in such theories). The important point in the
present context is that there exists a quantum mechanical relativistic setting
in which the S-matrix is Poincar\'{e} invariant, fulfills macro-causality and
implements interaction without using fields.

In this way one learns to appreciate the fundamental difference between
quantum theories which have no maximal velocity and those which have. As a
quantum mechanical theory DPI only leads to statistical "effective" finite
velocity propagation for asymptotically large time-like separations between
localized events as they occur in scattering theory. With other words the
causal propagation between Born-localized events is recovered only in the
sense of asymptotically large timelike distances. This explains in particular
why in such theories the S-matrix is Poincar\'{e} invariant. Saying that DPI
is macro- but not micro-causal implies that it cannot be used to study
properties of local propagation over finite distances. Asymptotically Fermi's
Gedankenexperiment leads to the desired result in QM (DPI) and QFT. But only
in the different notion of causal propagation which is totally characteristic
for QFT and does not exist in DPI. For relativistic scattering theory be it
DPI or QFT, the projectors and the related probablities which come with B-N-W
localization are indispensible. 

So at the root of the alleged QM-QFT (particle-field) antagonism is the
existence of two very different concepts of localization namely the
\textit{Born localization} which is the only localization for QM, and the
\textit{modular localization} which is the one underlying the causal locality
notion in QFT together with Born localization. the two only coalesce for
infinite timelike separations of events. For scattering theory in any
relativistic theory be it DPI or QFT one needs the Born-localization. In fact
one glance into the original paper by Born reveals that the probability
interpretation was made on scattering amplitudes leading to what is nowadays
called cross section in the Born approximation, the x-space wave functions; on
the other hand without modular localization there would be no
interaction-induced vacuum polarization  and instead the world at finite
distances would be filled with little acausal poltergeist-daemons.

Whereas QM only knows the Born localization, QFT requires both,
Born-localization for (the wave functions of) particles before and after a
scattering event, and modular localization\footnote{Modular localization is
the same as the causal localization inherent in QFT after one liberates the
letter from the contingencies of particular selected fields.} in connection
with fields and local observables\footnote{Particles are objects with a
well-defined ontological status, whereas (composite) fields form an infinite
set of coordinatizations which generate the local algebras. Modular
localization is the localization property which is independent of what field
coordinatization has been used.}. Without Born localization and the associated
projectors, there would be no scattering theory leading to cross sections and
QFT would become just a mathematical playground. 

In contradistinction to DPI, in interacting QFT there is no way in which in
the presence of interactions the notion of \textit{particles at finite times}
can be saved. The statement that an isolated relativistic particle cannot be
localized below its Compton wave length refers to the (Newton-Wigner
adaptation of the) Born localization and is meant, as all statements involving
Born localization, in an \textit{effective} probabilistic sense. Only in the
timelike asymptotic limit between two Born localization events, sharp
geometric relations with c being the maximal velocity emerge; this is
precisely what one needs to obtain a Poincar\'{e} invariant macrocausal
S-matrix. The maximal velocity in the sense of asymptotic expectations in
suitable states is of course the same mechanism which in nonrelativistic QM
leads to material-dependent acoustic velocities. 

The first part focusses on the radical difference between the Newton-Wigner
(NW) localization (the name for the Born localization after the adaptation to
the relativistic particle setting) and the localization which is inherent in
QFT, which in its intrinsic form, i.e. liberated from singular pointlike
"field coordinatizations", is referred to as \textit{modular localization
}\cite{BGL}\cite{Sch}\cite{MSY}\textit{. }The terminology\textit{\ }has its
origin in the fact that it is backed up by a mathematical theory within the
setting of operator algebras which bears the name
Tomita-Takesaki\footnote{Tomita was a Japanese mathematician who discovered
the main properties of the theory in the first half of the 60s, but it needed
a lot of polishing in order to be accepted by the mathematical community, and
this is where the name Takesaki entered.} \textit{modular theory.} Within the
\ setting of thermal QFT, physicists independently discovered various aspects
of this theory \cite{Haag}. Its relevance for causal localization was only
spotted a decade later \cite{Bi-Wi} and the appreciation of its use in
problems of thermal behavior at causal- and event- horizons and black hole
physics had to wait another decade \cite{Sew}.

The last subsection of the first part presents LQP as the result of
\textit{relative positioning} of a finite (and rather small) number of
\textit{monads} within a Hilbert space. This shows the enormous conceptual
distance between QM and LQP. Here we are using a terminology which Leibniz
introduced in a philosophical-ontological context. Whereas a single monad also
appears in different contexts e.g. the information theoretical interpretation
of bipartite spin algebras in suitable singular states \cite{Werner}%
\cite{Keyl-M}, the modular positioning of several copies is totally
characteristic for LQP. Although its physical and mathematical content is
quite different from Mermin's \cite{Mermin} new look (the
"Ithaca-interpretation" of QM) at quantum mechanical reality exclusively in
terms of correlations between subsystems, they share the aspect of
understanding reality in relational terms. Mathematically a monad in the sense
of this article is the unique hyperfinite type III$_{1}$ factor algebra to
which all local algebras in LQP are isomorphic, so all concrete monads are
copies of the abstract monad. Naturally a monade has no structure per se, the
reality emerges from their relation to each other.

Whereas for Newton physical reality consisted of matter moving in a fixed
space according to a universal time, reality for Leibniz emerges from
interrelations between monads with spacetime serving as ordering device. The
modular positioning of monads goes one step further in that even the Minkowski
spacetime together with its invariance group the Poincar\'{e} group appears as
a consequence of positioning in a more abstract sense namely of a finite
number of monads in a joint Hilbert space (subsection 7). For actual
constructions of interacting LQP models it is however advantageous to start
with one monad and the action of the Poincar\'{e} group on it. 

The algebraic structure of QM on the other hand, relativistic or not, has no
such monad structure; the global algebra as well as all Born-localized
subalgebras in ground states are always of type I i.e. either the algebra of
all bounded operators $B(H)$ in an appropriate Hilbert space or multiples
thereof. Correlations are characteristic features of quantum mechanical
states, whereas for the characterization of a QM system global operators as
the Hamiltonian are indispensable.

The second part addresses two important consequences of vacuum polarization,
the first subsection deals with \textit{localization entropy} and recalls its
area proportionality which is a more recent result \cite{S1}\cite{S2}. The
thermal aspects of localization have astrophysical and cosmological
consequences for black holes and for the cosmological constant problem which
will also be the subject of our discussion in that section. Our particular
interest is to look for an interface between QFT in CST and QG. Several issues
which in the past were expected to delimit the interface between QFT and QG,
including the two mentioned ones, are now believed to be taken care of within
QFT in CST extended by backreaction.

In particular some of the estimates of the cosmological constant which are
based on the filling up of energy levels, similar to the construction of the
Fermi surface in condensed matter physics, are already in trouble with the
QM/QFT interface. These estimates violates \textit{local covariance} (local
diffeomorphism equivalence) which as one of QFT in CST most cherished
principles is basically the locality principle of QFT extended with the
appropriately adapted local covariance from Einstein's classical theory. In
the title of one of Hollands and Wald's papers one finds the following advice
for avoiding such calculations: \textit{Quantum Field Theory Is Not Merely
Quantum Mechanics Applied to Low Energy Effective Degrees of Freedom}
\cite{Ho-Wa}. A model calculation without cutoff and in agreement with local
covariance and backreaction can be found in \cite{DFP}

\section{The interface between quantum mechanics and quantum field theory}

Shortly after the discovery of field quantization in the second half of the
1920s, there were two opposed viewpoints about its content and purpose
represented by Dirac and Jordan. Dirac's position was that quantum theory
should stand for \textit{quantizing a true classical reality}%
\footnote{Jordan's extreme formal positivistic point of view allowed him to
quantize everything which fitted into the classical Lagrangian field formalism
independent of whether it had a classical reality or not.} which meant field
quantization for electromagnetism and particle quantization for the massive
particles. Jordan, on the other hand proposed an uncompromising field
quantization point of view; all what can be quantized should be quantized,
independent of whether there is a classical reality or not. The more radical
field quantization finally won the argument, but ironically it was Dirac's
particle setting (the hole theory) and not Jordan's application of Murphy's
law to all field objects which contributed the richest structural property to
QFT, namely antiparticles/anticharges. It was also the hole theory in which
the first perturbative QED computations (which entered the textbooks of
Heitler and Wenzel) were done, before it was recognized that this setting was
not really consistent. This inconsistency \ showed up in problems involving
renormalization in which \textit{vacuum polarization} plays the essential
role. The successful perturbative renormalization of QED was also the end of
hole theory and the beginning of Dirac's late conversion to QFT as the general
setting for relativistic particle physics at the beginning of the 50s.

Vacuum polarization is a very peculiar phenomenon which in the special context
of currents and the associated local charges of a complex free Bose field was
noticed already in the 30s by Heisenberg \cite{Hei}. But only when Furry and
Oppenheimer \cite{Fu-Op} studied perturbative interactions of Lagrangian
fields and noticed to their amazement that the Lagrangian field applied to the
vacuum created inevitably some additional particle-antiparticle pairs in
addition to the expected one-particle state. The number of these pairs
increase with the perturbative order, pointing towards the fact that one has
to deal with infinite polarization clouds in case of sharp localization.
Whenever one tries in an interacting theory to create particles via local
disturbances of the vacuum these vacuum polarization clouds corrupt precisely
those particles which one intends to create. In the presence of interactions
the notion of particles in local regions is simply meaningless; they only
appear in the form of incoming and outgoing asymptotic particle configurations.

In the next subsection it will be shown that relativistic QM in the form of
DPI can indeed be consistently formulated and this setting can even be
extended to incorporate creation and annihilation channels \cite{P}. This goes
along way to vindicate Dirac's relativistic particle viewpoint. But it does
not vindicate it completely since theories which start as particle theories
but then lead to vacuum polarization as Dirac's hole theory are at the end
inconsistent. this indeed possible. Although the DPI setting will only be
formulated for elastic scattering processes, it can be extended by adding
creation channels

This essay does not attempt to advertise DPI as an alternative particle
description to QFT, it is only meant as a conceptual challenge. By contrasting
the latter with the former one learn to appreciate the conceptual depth of QFT
and one becomes aware of its still unexplored regions. DPI is basically a
relativistic \textit{particle} setting i.e. it deals only with properties
which can be formulated in terms of particles; this limits causality
properties to macro-causality i.e. spacelike cluster factorization and
timelike causal rescattering. This setting is as well understood as QM; one
would be surprised to find still unilluminated regions. \ In contrast nobody
who has studied QFT beyond a textbook level would claim to know what those
postulates or axioms by which one tries to define QFT really lead to. Even
now, 80 years after its discovery, one is deeply impressed that something that
old can still reveal secrets. The last subsection of the present section
illustrates this point by an interesting recent example.

\subsection{Direct particle interactions, relativistic QM}

The Coester-Polyzou \ theory of \textit{direct particle interactions} (DPI),
where direct means not field-mediated, is a relativistic theory in the sense
of representation theory of the Poincar\'{e} group which among other things
leads to a Poincar\'{e} invariant S-matrix. Every property which can be
formulated in terms of particles, as the cluster factorization into systems
with a lesser number of particles and other aspects of macrocausality, is
fulfilled in this setting. The S-matrix does not fulfill analyticity
properties as the crossing property whose derivation relies on the existence
of local interpolating fields.

In contradistinction to the more fundamental locally covariant QFT, DPI is
only a phenomenological setting, but one which is consistent with every
property which can be expressed in terms of relativistic particles only. For a
long time it was only known how to deal with \textit{two} interacting
particles. In that case one goes to the c. m. system and modifies the
invariant energy operator. Assuming for simplicity identical scalar Bosons,
the c.m. invariant energy operator is $2\sqrt{p^{2}+m^{2}}~$and the
interaction is introduced by adding an interaction term $v$%

\begin{equation}
M=2\sqrt{\vec{p}^{2}+m^{2}}+v,~~H=\sqrt{\vec{P}^{2}+M^{2}}%
\end{equation}
where the invariant potential $v$ depends on the c.m. variables $p,q$ in an
invariant manner i.e. such that $M$ commutes with the Poincar\'{e} generators
of the 2-particle system which is a tensor product of two one-particle systems.

One may follow Bakamjian and Thomas (BT) \cite{BT} and choose the Poincar\'{e}
generators in their way the interaction does not affect them directly apart
from the Hamiltonian. Denoting the interaction-free generators by a subscript
$0$ one arrives at the following system of two-particle generators%
\begin{align}
\vec{K} &  =\frac{1}{2}(\vec{X}_{0}H+H\vec{X}_{0})-\vec{J}\times\vec{P}%
_{0}(M+H)^{-1}\\
\vec{J} &  =\vec{J}_{0}-\vec{X}_{0}\times\vec{P}_{0}\nonumber
\end{align}

The interaction $v$ may be taken as a \textit{local} function in the relative
coordinate which is conjugate to the relative momentum p in the c.m. system;
but since the scheme anyhow does not lead to local differential equations,
there is not much to be gained from such a choice. The Wigner canonical spin
$\vec{J}_{0}$ commutes with $\vec{P}=\vec{P}_{0}$ and $\vec{X}=\vec{X}_{.0}$
and is related to the Pauli-Lubanski vector $W_{\mu}=\varepsilon_{\mu\nu
\kappa\lambda}P^{\nu}M^{\kappa\lambda}$ .

As in the nonrelativistic setting, short ranged interactions $v$ lead to
M\o ller operators and S-matrices via a converging sequence of unitaries
formed from the free and interacting Hamiltonian%
\begin{align}
\Omega_{\pm}(H,H_{0}) &  =\lim_{t\rightarrow\pm\infty}e^{iHt}e^{-H_{0}t}\\
\Omega_{\pm}(M,M_{0}) &  =\Omega_{\pm}(H,H_{0})\nonumber\\
S &  =\Omega_{+}^{\ast}\Omega_{-}\nonumber
\end{align}
The identity in the second line is the consequence of a theorem which say that
the limit is not affected if instead of $M$ we take a positive function of $M$
as $H(M),$ as long as $H_{0}$ is the same function of $M_{0}.$ This insures
the \textit{frame-independence of the M\o ller operators and the S-matrix}.
Apart from this identity for operators and their positive functions, which is
not needed in the nonrelativistic scattering, the rest behaves just as in
nonrelativistic scattering theory. As in standard QM, the 2-particle cluster
property is the statement that $\Omega_{\pm}^{(2)}\rightarrow\mathbf{1,}$
$S^{(2)}\rightarrow\mathbf{1,}$ i.e. the scattering formalism is identical. In
particular the two particle cluster property, which says that for short range
interactions the S-matrix approaches the identity if one separates the center
of the wave packets of the two incoming particles, holds also for the
relativistic case.

The implementation of clustering is much more delicate for 3 particles as can
be seen from the fact that the first attempts were started in 1965 by Coester
\cite{Coe} and considerably later generalized (in collaboration with Polyzou
\cite{C-P}) to arbitrary high particle number. To anticipate the result below,
DPI leads to a consistent scheme which fulfills cluster factorization but it
has no useful second quantized formulation so it may stand accused of lack of
elegance, and since we are inclined to view less elegant theories also as less
fundamental, we would not trade this phenomenological relativistic theory
(arbitrary potential functions instead of pontlike coupling parameters) for
QFT. It is also more nonlocal and nonlinear than QM, This had to be expected
since adding particles does not mean adding terms to the Hamiltonian as in
Schroedinger QM.

The BT form for the generators can be achieved inductively for an arbitrary
number of particles. As will be seen, the advantage of this form is that in
passing from n-1 to n-particles the interactions simply add and one ends up
with Poincar\'{e} group generators for an interacting n-particle system. But
for $n>2$ the aforementioned subtle problem with the cluster property arises;
whereas this iterative construction in the nonrelativistic setting complies
with cluster separability, this is not the case in the relativistic context.

This problem shows up for the first time in the presence of 3 particles
\cite{Coe}. The BT iteration from 2 to 3 particles gives the 3-particle mass operator%

\begin{align}
M  &  =M_{0}+V_{12}+V_{13}+V_{23}+V_{123}\label{add}\\
V_{12}  &  =M(12,3)-M_{0}(12;3),~M(12,3)=\sqrt{\vec{p}_{12,3}^{2}+M_{12}^{2}%
}+\sqrt{\vec{p}_{12,3}^{2}+m^{2}}\nonumber
\end{align}
and the $M(ij,k)$ result from cyclic permutations. Here $M(12,3)$ denotes the
3-particle invariant mass in case the third particle is a \textquotedblleft
spectator\textquotedblright,\ which by definition does not interact with 1 and
2. The momentum in the last line is the relative momentum between the
$(12)$-cluster and particle $3$ in the joint c.m. system and $M_{12}$ is the
associated two-particle mass i.e. the invariant energy in the $(12)$ c.m system.

As in the nonrelativistic case, one can always add a totally connected
contribution. Setting this contribution to zero, the 3-particle mass operator
only depends on the two-particle interaction $v.~$But contrary to the
nonrelativistic case, the BT generators constructed with $M$ do not fulfill
the cluster separability requirement as it stands. The latter demands that if
the interaction between two clusters is removed, the unitary representation
factorizes into that of the product of the two clusters.

One expects that shifting the third particle to infinity will render it a
spectator and result in a factorization $U_{12,3}\rightarrow U_{12}\otimes
U_{3}$. Unfortunately what really happens is that the $(12)$ interaction also
gets switched off i.e. $U_{123}\rightarrow U_{1}\otimes U_{2}\otimes U_{3}$ .
The reason for this violation of the cluster separability property, as a
simple calculation using the transformation formula from c.m. variables to the
original $p_{i}$, i = 1, 2, 3 shows \cite{C-P}, is that although the spatial
translation in the original system (instead of the $12,3$ c.m. system) does
remove the third particle to infinity as it should, unfortunately it also
drives the two-particle mass operator (with which it does not commute) towards
its free value which violates clustering.

In other words the BT produces a Poincar\'{e} covariant 3-particle interaction
which is additive in the respective c.m. interaction terms (\ref{add}), but
the Poincar\'{e} representation $U$ of the resulting system will not be
cluster-separable. However, as shown first in \cite{Coe}, at least the
3-particle S-matrix computed in the additive BT scheme turns out to have the
cluster factorization property. But without implementing the correct cluster
factorization not only for the S-matrix but also for the 3-particle
Poincar\'{e} generators there is no chance to proceed to a clustering
4-particle S-matrix.

Fortunately there always exist unitaries which transform BT systems into
cluster-separable systems \textit{without affecting the S-matrix}. Such
transformations are called \textit{scattering equivalences. }They were first
introduced into QM by Sokolov \cite{So} and their intuitive content is related
to a certain insensitivity of the scattering operator under quasilocal changes
of the quantum mechanical description at finite times. This is reminiscent of
the insensitivity of the S-matrix in QFT against local changes in the
interpolating field-coordinatizations\footnote{In field theoretic terminology
this means changing the pointlike field by passing to another (composite)
field in the same equivalence class (Borchers class) or in the setting of AQFT
by picking another operator from a local operator algebra.} by e.g. using
composites instead of the Lagrangian field. The notion of scattering
equivalences is conveniently described in terms of a subalgebra of
\textit{asymptotically constant operators} $C$ defined by%
\begin{align}
\lim_{t\rightarrow\pm\infty}C^{\#}e^{iH_{0}t}\psi &  =0\\
\lim_{t\rightarrow\pm\infty}\left(  V^{\#}-1\right)  e^{iH_{0}t}\psi &
=0\nonumber
\end{align}
where $C^{\#}$ stands for both $C$ and $C^{\ast}$. These operators, which
vanish on dissipating free wave packets in configuration space, form a
*-subalgebra which extends naturally to a $C^{\ast}$-algebra $\mathcal{C}$. A
scattering equivalence is a unitary member $V\in$ $\mathcal{C}$ which is
asymptotically equal to the identity (the content of the second line).
Applying this asymptotic equivalence relation to the M\o ller operator one obtains%

\begin{equation}
\Omega_{\pm}(VHV^{\ast},VH_{0}V^{\ast})=V\Omega_{\pm}(H,H_{0})
\end{equation}
so that the $V$ cancels out in the S-matrix. Scattering equivalences do
however change the interacting representations of the Poincar\'{e} group
according to $U(\Lambda,a)\rightarrow VU(\Lambda,a)V^{\ast}.$

The upshot is that there exists a clustering Hamiltonian $H_{clu}$ which is
unitarily related to the BT Hamiltonian $H_{BT}$ i.e. $H_{clu}=BH_{BT}B^{\ast
}$ such that $B\in\mathcal{C}.~$is uniquely determined in terms of the
scattering data computed from $H_{BT}.$ It is precisely this clustering of
$H_{clu}$ which is needed for obtaining a clustering 4-particle S-matrix which
is cluster-associated with the $S^{(3)}$. With the help of $M_{clu}$ one
defines a 4-particle interaction following the additive BT prescription; the
subsequent scattering formalism leads to a clustering 4-particle S-matrix and
again one would not be able to go to n=5 without passing from the BT to the
cluster-factorizing 4-particle Poincar\'{e} group representation. Coester and
Polyzou showed \cite{C-P} that this procedure can be iterated and hence one
arrives at the following statement

\textbf{Statement}: \textit{The freedom of choosing scattering equivalences
can be used to convert the Bakamijan-Thomas presentation of multi-particle
Poincar\'{e} generators into a cluster-factorizing representation. In this way
a cluster-factorizing S-matrix }$S^{(n)}$\textit{ associated to a BT
representation }$H_{BT}$\textit{ (in which clustering mass operator }%
$M_{clu}^{(n-1)}$\textit{ was used) leads via the construction of }%
$M_{clu}^{(n)}$\textit{ to a S-matrix }$S^{(n+1)}$\textit{ which clusters in
terms of all the previously determined }$S^{(k)},k<n.$ \textit{The use of
scattering equivalences impedes the existence of a 2}$^{nd}$\textit{ quantized
formalism.}

For a proof we refer to the original papers \cite{C-P}\cite{P}. In passing we
mention that the minimal extension, i.e. the one determined uniquely in terms
of the two-particle interaction $v$) from n to n+1 for $n>3,$ contains
\textit{connected 3-and higher particle interactions} which are nonlinear
expressions (involving nested roots) in terms of the original two-particle
$v.~$This is another unexpected phenomenon as compared to the nonrelativistic case.

This theorem shows that it is possible to construct a relativistic theory
which only uses particle concepts only, thus correcting an old folklore which
says relativity + clustering = QFT. Whether one should call this DPI theory
"relativistic QM" is a matter of taste, it depends on what significance one
attributes to those unusual scattering equivalences. But in any case it
defines a \textit{relativistic S-matrix setting }with the correct particle
behavior\textit{. }In this context one should also mention that the S-matrix
bootstrap approach never addressed these macro-causality problems and this
also holds for its heir the contemporary string theory.

As mentioned above Coester and Polyzou also showed that this relativistic
setting can be extended to processes which maintain cluster factorization in
the presence of a finite number of creation/annihilation channels, showing, as
mentioned before, that the mere presence of particle creation is not
characteristic for QFT\footnote{It does not look very likely that the S-matrix
of QFT can be approximated as a limit of DPI with particle creation.}.
Different from the nonrelativistic Schroedinger QM, the superselection rule
for masses of particles which results from Galilei invariance for
nonrelativistic QM does not carry over to the relativistic setting; in this
respect DPI is less restrictive than its Galilei-invariant QM counterpart
where such creation processes are forbidden.

Certain properties which are automatic consequences of locality in QFT but can
be formulated solely in terms of particles as TCP symmetry, the existence of
anti-particles, the spin-statistics connection, can be added "by hand". Other
properties which are on-shell relics of locality which QFT imprints on the
S-matrix and which require the notion of analytic continuation in on-shell
particle momenta, as e.g. the crossing property, cannot be implemented in the
QM setting of DPI.

\subsection{First brush with the intricacies of the particles-field problems
in QFT}

Interacting QFT in contrast to QM (Schr\"{o}dinger-QM or relativistic DPI),
does not admit a particle interpretation at finite times\footnote{Although the
one-particle states and their multiparticle counterparts are global states in
the Hilbert space, they are not accessible by acting locally on the vacuum.
Scattering theory is the only known nonlocal intervention.}. If it would not
be for the asymptotic scattering interpretation in terms of incoming/outgoing
particles associated with the free in/out fields, there would be hardly
anything of a non-fleeting nature which can be measured. In QFT in CST and
thermal QFT where this particle concept is missing, the set of conceivable
measurements is ostensibly meagre and is essentially reduced to energy- and
entropy- densities as in thermal systems and black hole radiation.

Since the notion of particle is often used in a more general sense than in
this paper, it may be helpful to have an interlude on this topic. By particle
I mean an asymptotically stable object which forms the tensor product basis
for an asymptotically complete description. It is precisely the particle
concept which furnishes QFT with a (LSZ, Haag-Ruelle) complete asymptotic
particle interpretation\footnote{The asymptotic completeness property was for
the first time established (together with a recent existence proof) in a
family of factorizing two-dimensional models (see the section on modular
localization) with nontrivial scattering.}, so that a Fock space tensor
structure is imposed on the Hilbert space of the interacting system. The
physics behind it is the idea that if we were to cobble the asymptotic
spacetime region with counters and monitor coincidences of localization
events, then the n-fold coincidence/anticoincidence (the latter in order to
insure that we caught all particles) set up would eventually remain stable
because the far removed localization centers would have ceased to interact and
from there on move freely i.e. one would be in a region where the
Newton-Wigner adaptation of the Born position operator would lead to genuinely
Poincar\'{e} invariant transition probabilities.

The particle concept in QFT is therefore precisely applicable where it is
needed, namely for asymptotically separated Born-localized events; thus the
invariant S-matrix has no memory about the reference-system-dependent Born
localization and the question of what particle counters really count in finite
regions becomes academic. In fact the careless use of the B-N-W localization
for finite distances is known to lead to unphysical superluminal effects; in
that case one should formulate the problem in the setting of the modular localization.

Tying up the particle concept in QFT to asymptotically stable
counter-coincidences can be traced back to a seminal paper by Haag and Swieca
\cite{Ha-Sw}. In that paper it was noticed for the first time that the phase
space volume in QFT unlike that in QM is not finite but its cardinality is
very mild (the phase space is \textit{nuclear}). This is yet another line of
unexpected different consequences \cite{Swieca} resulting from the different
localization concepts in QM and QFT, but this interested topic will not be
pursued here.

Not all particles comply with this definition; in fact all electrically
charged particles are \textit{infraparticles} i.e. objects which are
asymptotically stable up to an unobserved cloud of infinitely may infrared
photons whose presence has the consequence that instead of the mass shell
$p^{2}=m^{2}$ the mass $m$ of the charged particle only denotes the start of
cut instead of a mass shell delta function \cite{infra}. The naive scattering
theory leads to infrared divergencies which cannot be cured by renormalizing
parameters, but rather requires a significant change of scattering theory.
Since for the problems at hand this is of less importance we leave it at these remarks.

It is the \textit{asymptotic particle structure} which leads to the
observational richness of QFT. Once we leave this setting by going to curved
spacetime or to QFT in KMS thermal states representations, or if we restrict a
Minkowski spacetime theory to a Rindler wedge with the Hamiltonian being now
the boost operator with its two-sided spectrum, we are loosing the setting of
scattering theory of particles and its observational wealth. The restriction
to the Rindler world preserves the Fock space particle structure of the free
field Minkowski QFT, but it looses its direct intrinsic significance with
respect to the Rindler situation\footnote{There is of course the mathematical
possibility of choosing a groundstate representation for a Rindler world
instead of restricting the Minkowski vacuum. In that case it is not clear
whether in the presence of interactions the exitations above this ground state
have the Haag-Swieca asymptotic localization stability i.e. whether scattering
theory applies to such a situation. It would be interesting to (dis)prove the
validity of Haag-Ruelle scattering theory in such a situation.}. In the
Rindler world since the Minkowski vacuum is now a thermal state and there is
no particle scattering theory in the "boost time" in such a thermal situation.

Of course there remains the possibility to measure thermal excitations in an
\textit{Unruh counter }\cite{Un} or to use a counter to register Hawking
\cite{Haw} radiation. or to determine the energy density in a cosmological
reference state (see also last section). In this case one is not measuring
individual particles but rather an energy density. An famous example for a
kind of measurement with great physical significance for the development of
cosmology is the cosmic background radiation which is the expectation of the
energy density in the cosmic reference state. In such situations one does not
only loose the Poincar\'{e} symmetry but together with it the vacuum as well
as the particle state.

This raises the question whether the result of over 60 years accumulated
knowledge about scattering of particles still find a conceptual place in the
more general QFT in CST or whether the only data consistent CST are those
obtained placing a counters into a cosmic reference state and measuring
expectations of the energy. On one extreme end are those who claim that
particles have no place in QFT in CST, one is rather forced to abandon
particles altogether and adopt the point that one measures fields, in
particular the energy density in the cosmic state. This point of view one
finds in particular in recent publications of Wald \cite{Wald2}. The only kind
of field for which one can envisage a field measurements without thinking in
terms of particles is the electromagnetic field apart from this exception all
other fields serve a interpolating fields for particles. So the question what
fields interpolate in the setting of CST, if not particles, remains open.

Formally the local covariance principle forces the construction of a QFT on
all causally complete manifolds and their submanifolds at once. So the QFT in
Minkowski spacetime with its particle interpretation is always part of the
solution. What one would like to have is a more direct physical connection
e.g. a particle concept in the tangent space or something in this direction.

The conceptual differences between a DPI relativistic QM and QFT are enormous,
but in order to appreciate this, one has to become acquainted with structural
properties of QFT which are somewhat removed from the standard properties of
the Lagrangian setting and therefore have not entered textbooks; it is the
main purpose of the following sections to highlight these contrasts by going
more deeply into QFT.

There are certain folkloric statements about the relation QM--QFT whose
dismissal does not require much conceptual sophistication. For example in
trying to make QFT more susceptive to newcomers it is sometimes said that a
free field is nothing more than a collection of infinitely many coupled
oscillators. Although not outright wrong, this characterization misses the
most important property of how spacetime enters as an ordering principle into
QFT. It would not help any newcomer who knows the quantum oscillator, but has
not met a free field before, to construct a free field from such a verbal
description. Even if he manages to write down the formula of the free field he
would still have to appreciate that the most important aspect is the causal
localization and not that what oscillates. This is somewhat reminiscent of the
alleged virtue from equating QM via Schr\"{o}dinger's formulation with
classical wave theory. What may be gained for a newcomer by appealing to his
computational abilities acquired in classical electrodynamics, is more than
lost in the conceptual problems which he confronts later when facing the
subtleties of quantum physics.

\subsection{Quantum mechanical Born localization versus covariant localization
in LQP}

Let us know come to the main point namely the difference between QM and LQP in
terms of their localization concept. As it should be clear from previous
remarks we will use the word \textit{Born localization} for the probability
density of the x-space Schroedinger wave function $p(x)=\left\vert
\psi(x)\right\vert ^{2};$ the adaptation to the invariant inner product of
relativistic wave functions was done by Newton and Wigner \cite{N-W} and will
be continued to be referred to as B-N-W localization. Being a bona fide
probability density, one may characterize the Born localization in a spatial
region $R\in\mathbb{R}^{3}$ at a given time in terms of a \ localization
projector $P(R).$ The standard version of QM and the various settings of
measurement theory rely heavily on these projectors. Without Born localization
and the ensuing projectors it would be impossible to formulate the conceptual
basis for the time-dependent scattering theory.

The B-N-W position operator and its family of spatial region-dependent
projectors are not covariant under Lorentz boosts. For Wigner, who was not
aware of the existence of the covariant modular localization, this frame
dependence raised doubts about the conceptual soundness of QFT. Apparently the
existence of completely covariant correlation functions in renormalized
perturbation theory did not satisfy him, he wanted an understanding from first principles.

The fact that modular localization remained closed to Wigner may be seen as an
indication of its subtlety; the standard operator formalism as used in QM
contains not the slightest hint in its direction.

The lack of covariance of B-N-W localization in finite time propagation leads
to frame-dependence and superluminal contributions, which is why the
terminology "relativistic QM" has to be taken with a grain of salt. However,
as already emphasized, in the asymptotic limit of large timelike separation as
required in scattering theory, the covariance, frame-independence and causal
relations are recovered. With other words one obtains a Poincar\'{e}-invariant
unitary S-matrix whose DPI construction can also be shown to guaranty also the
validity of all the macro-causality requirements (spacelike clustering,
absence of timelike precursors, causal rescattering) which can be formulated
in a particle setting without taking recourse to interpolating local fields.
Even though the localizations of the individual particles are frame-dependent,
the asymptotic relation between N-W-localized events is given in terms of the
geometrically associated \textit{covariant on-shell momenta} or 4-velocities.
In fact all observations on particles always involve B-N-W localization measurements.

The situation of propagation of DPI is similar to that of propagation of
acoustic waves in an elastic medium; although in neither case there is a
limiting velocity there exists a maximal "effective" velocity, for DPI this is
c and in the acoustic case this is the velocity of sound.

In comparing QM with QFT it is often convenient in discussions about
conceptual issues to rephrase the content of (nonrelativistic) QM in terms of
operator algebras and states in the sense of positive expectation functional
on operator algebras; in this way one also achieves more similarity with the
formalism of QFT where this abstraction becomes important. In this Fock space
setting the basic operators are creation/annihilation operators $a^{\#}%
(\mathbf{x})$ with%
\begin{equation}
\left[  a(\mathbf{x}),a^{\ast}(\mathbf{y})\right]  _{grad}=\delta
(\mathbf{x}-\mathbf{y})\label{qu}%
\end{equation}
where for Fermions the graded commutator is the anticommutator. The ground
state for T=0 zero matter density states is annihilated by $a(\mathbf{x})$
whereas for fine density one has a filled below the Fermi surface state for
Fermions and a Bose-Einstein condensate for Bosons. In QFT the identification
of pure states with state-vectors of a Hilbert space has no intrinsic meaning
and often cannot be maintained in concrete situations. For the same reasons of
achieving a unified description we use the multi-particle (Fock space) setting
instead of the Schroedinger formulation. Although DPI is formulated in Fock
space there is no second quantized formalism (\ref{qu}), which renders the
formalism less elegant and more detailed than its nonrelativistic counterpart.

The global algebra which contains all observables independent of their
localization is the algebra $B(H)$ of all bounded operators in Hilbert space.
Physically important unbounded operators are not members but rather have the
mathematical status of being affiliated with $B(H)$ and its subalgebras; this
bookkeeping makes it possible to apply powerful theorems from the theory of
operator algebras (whereas unbounded operators are treated on a case to case
basis). $B(H)$ is the correct global description whenever the physical system
under discussion arises as the weak closure of a ground state representation
of an irreducible system of operators\footnote{The closure in a thermal
equilibrium state associated with a continuous spectrum Hamiltonian leads to a
unitarily inequivalent (type III) operator algebra without minimal
projectors.} be it QM or LQP$.$ According to the classification of operator
algebras, $B(H)$ and all its multiples are of Murray von Neumann type
$I_{\infty}$ whose characteristic property is the existence of minimal
projectors; in the irreducible case these are the one-dimensional projectors
belonging to measurements which cannot be refined$.$ There are prominent
physical states which lead to different global situations as e.g. thermal KMS
states but for the time being our interest is in ground states.

The structural differences between QM and LQP emerge as soon as one uses
localization in order to provide a physical substructure to $B(H)$. It is well
known that a dissection of space into nonoverlapping spatial regions i.e.
$\mathbb{R}^{3}=\cup_{i}R_{i}$ implies via Born localization a tensor
factorization of $B(H)$ and $H$%
\begin{align}
&  B(H)=%
{\displaystyle\bigotimes\limits_{i}}
B(H(R_{i}))\\
&  H=%
{\displaystyle\bigotimes_{i}}
H(R_{i}),~P(R_{i})H=H(R_{i})\nonumber\\
&  \mathbf{\vec{X}}_{op}=\int a^{\ast}(\vec{x})\vec{x}a(\vec{x})d^{3}%
x=\int\vec{x}dP(\vec{x})
\end{align}
where the third line contains the definition of the position operator and its
spectral decomposition in the bosonic Fock space. Hence there is orthogonality
between subspaces belonging to localizations in nonoverlapping regions
(orthogonal Born projectors) and one may talk about states which are pure in
$H(R_{i}).$ A pure state in the global algebra $B(H)$ may not be of the tensor
product form but may rather describe a superposition of factorizing states;
the Schmidt decomposition is a method to achieve this with an intrinsically
determined basis in the two factors. States which are not tensor products but
rather superpositions of such are called entangled and their reduced density
matrix obtained by averaging outside a region $R_{i}$ describes a mixed state
on $B(H(R_{i}))$. This is the standard formulation of QM in which pure states
are vectors and mixed states are density matrices.

Although this quantum mechanical entanglement can be related with the notion
of entropy, it is an entropy in the sense of \textit{information theory} and
not in the thermal sense of thermodynamics, i.e. one cannot create a
temperature as a quantitative measure of the degree of quantum mechanical
entanglement which results from Born-restricting pure global states to a
finite region and its outside environment. The net structure of $B(H)$ in
terms of the $B(H(R_{i})$) is of a kinematical kind, it does not create a new
Hamiltonian with respect to which the reduced state becomes a KMS state. The
quantum mechanical dynamics through a Hamiltonian shows that the tensor
factorization from Born localization at one time is almost instantaneously
lost in the time-development, as expected of a theory of without a maximal
propagation speed.

The LQP counterpart of the Born-localized subalgebras at a fixed time are the
observable algebras $\mathcal{A(O)}$ for causally completed ($\mathcal{O=O}%
^{\prime\prime},$ the causal complement taken twice) spacetime regions
$\mathcal{O}$; they form what is called in the terminology of LQP a
\textit{local net} \{$\mathcal{A(O)}$\}$_{\mathcal{O}\subset M}$ of operator
algebras indexed by regions in Minkowski spacetime $\cup\mathcal{O=M}$ which
is subject to the natural and obvious requirements of isotony ($\mathcal{A(O}%
_{1}\mathcal{)}\subset\mathcal{A(O}_{2}\mathcal{)}$ if $O_{1}\subset O_{2}$)
and causal locality, i.e. the algebras commute for spacelike separated regions.

The connection with the standard formulation of QFT in terms of pointlike
fields is that smeared fields $\Phi(f)=\int\Phi(x)f(x)d^{4}x$ with
$suppf\subset\mathcal{O}~$under reasonable general conditions generate local
algebras. Pointlike fields, which by themselves are too singular to be
operators (even if admitting unboundedness), have a well-defined mathematical
meaning as operator-valued distributions. But as mentioned before, there are
myriads of fields which generate the same net of local operator algebras,
hence they play a similar role in LQP as coordinates in modern differential
geometry i.e. they coordinatize the net of spacetime indexed operator algebras
and only the latter has an intrinsic meaning. But as the use of particular
coordinates often facilitates geometrical calculations, the use of particular
fields with e.g. the lowest short-distance dimension within the infinite
charge equivalence class of fields can greatly simplify algebraic calculations
in QFT. Therefore it is a problem of practical importance to construct a
covariant basis of locally covariant pointlike fields of an equivalence class.

For massive free fields and for massless free fields of finite helicity such a
basis is especially simple; the "Wick-basis" of composite fields still follows
in part the logic of classical composites. This remains so even in the
presence of interactions in which case the Wick-ordering gets replaced by the
technically more demanding "normal ordering". For free fields in CST and the
definition of their composites it is important to require the \textit{local
covariant transformation behavior} under local isometries \cite{Ho-Wa2}. The
conceptual framework in the presence of interactions has also been understood
\cite{BFV}.

We now return to the main question namely what changes if we pass from the
Born localization of QM to the causal localization of LQP? The crucial
property is that a localized algebra $\mathcal{A(O)}\subset B(H)$ together
with its commutant $\mathcal{A(O)}^{\prime}$ (which under very general
conditions\footnote{In fact this duality relation can always be achieved by a
process of maximalization (Haag dualization) which increases the degrees of
freedom inside $\mathcal{O}.$ A pedagogical illustration based on the
"generalized free field" can be found in \cite{Du-Re}.} is equal to algebra of
the causal disjoint of $\mathcal{O}$ i.e. $\mathcal{A(O)}^{\prime
}=\mathcal{A(O}^{\prime}\mathcal{))}$ are two von Neumann factor algebras i.e.%
\begin{equation}
B(H)=\mathcal{A(O)\vee A(O)}^{\prime},\text{ }\mathcal{A(O)\cap A(O)}^{\prime
}=\mathbb{C}\mathbf{1}\label{loc}%
\end{equation}
In contrast to the QM algebras the local factor algebras are not of type I and
$B(H)$ does \textit{not tensor-factorize} in terms of them, in fact they
cannot even be embedded into a $B(H_{1})\otimes B(H_{2})$ tensor product. The
prize to pay for ignoring this important fact and imposing wrong structures is
the appearance of spurious ultraviolet divergences. On the positive side, as
will be seen later, without this significant change in the nature of algebras
there would be no holography onto causal horizons, no thermal behavior caused
by localization and a fortiori no area-proportional localization entropy.

In QM a pure state vector, which, with respect to a distinguished tensor
product basis in $H(R)\otimes H$($R\backslash\mathbb{R}^{3}$), is a nontrivial
superposition of tensor-basis states, will be generally become impure state if
restricted to $B(H(R))$; in the standard formalism (where only pure states are
represented by vectors) it is described by a density matrix. This phenomenon
of \textit{entanglement} is best described by the \textit{information
theoretic notion of entropy}. On the other hand each pure state on $B(H(R))$
or $B(H(R\backslash\mathbb{R}^{3}))$ originates from a pure state on $B(H).$

The situation in LQP is radically different since the local algebras as
$\mathcal{A(O})$ have \textit{no pure states at all}; so the dichotomy between
pure and mixed states breaks down and the kind of entanglement caused by field
theoretic localization is much more violent then that coming from
Born-localization\footnote{By introducing in addtion to free fields $A(x)$
which are covariant Fourier transforms also noncovariant Fourier transforms
$a(\vec{x},t),a^{\ast}(\vec{x},t)$ one can explicitly that they are relatively
nonlocal.} (see below). Unlike Born localization, causal localization is not
related to position operators and projectors $P(R);$ rather the operator
algebras $\mathcal{A(O})$ are of an entirely different kind than those met in
ground state QM (zero temperature); they are all isomorphic to one abstract
object, the hyperfinite type III$_{1}$ von Neumann factor also referred to as
\textit{the monad.} As will be seen later LQP creates its wealth from just
this one kind of brick; all the structural richness comes from positioning the
bricks, there is nothing in the bricks themselves. In a later section it will
later be explained how this emerges from modular localization and a related
operator formalism.

The situation does not change if one takes for $\mathcal{O}$ a region $R$ at a
fixed time; in fact in a theory with finite propagation one has$\mathcal{A}%
(R)=\mathcal{A}(D(R))$ where $D(R)$ is the diamond shaped double cone
subtended by $R$ (the causal shadow of $R$)$.$ Even if there are no pointlike
generators and if the theory only admits a macroscopically localized net of
algebras (e.g. a net of non-trivial wedge-localized factor algebras
$\mathcal{A}(W)$ which leads to trivial double cone intersection algebras
$\mathcal{A(O})),$ the algebras would still not tensor factorize i.e.
$B(H)\neq\mathcal{A}(W)\otimes\mathcal{A}(W^{\prime}).$ Hence the properties
under discussion are not directly related to the presence of singular
generating fields but are connected to the existence of well-defined causal
shadows. It turns out that there is a hidden singular aspect in the sharpness
of the $\mathcal{O}$-localization which generates infinitely large vacuum
polarization clouds on the causal horizon of the localization. In a later
section a method (splitting) will be presented which permits to define a
split-distance dependent but otherwise intrinsically defined finite thermal entropy.

Many divergencies in QFT are the result of conceptual errors in the
formulation resulting from tacitly identifying QFT with some sort of
relativistic QM\footnote{The correct treatment of perturbation theory which
takes into account the singular nature of pointlike quantum fields may yield
more free parameters than in the classical setting, but one is never required
to confront infinities or cut-offs.}, especially computations which ignore the
singular nature of pointlike localized fields. Conceptual mistakes are
facilitated by the fact that even nonlocal but covariant objects are singular;
this is evident from the Kallen-Lehmann representation of a covariant scalar
object%
\begin{equation}
\left\langle A(x)A(y)\right\rangle =\int\Delta_{+}(x-y,\kappa^{2})\rho
(\kappa^{2})d\kappa^{2}%
\end{equation}
which was proposed precisely to show that even without demanding locality, but
retaining only covariance and the Hilbert space structure (positivity), a
certain singular behavior of covariant objects is unavoidable. In the DPI
scheme this was avoided because covariance there is limited to asymptotic relations.

In the algebraic formulation the covariance requirement refers to the geometry
of the localization region $\mathcal{A}(\mathcal{O})$ i.e.%
\begin{equation}
U(a,\Lambda)\mathcal{A(O})U(a,\Lambda)^{\ast}=\mathcal{A}(\mathcal{O}%
_{a,\Lambda})
\end{equation}
whereas no additional requirement about the transformation behavior under
finite (tensor, spinor) Lorentz representations (which would bring back the
unboundedness and thus prevent the use of powerful theorems in operator
algebras) is imposed for the individual operators. The singular nature of
pointlike generators (if they exist) is then a purely mathematical
consequence. Using such singular objects in pointlike interactions in the same
way as one uses operators in QM leads to self-inflicted divergence problems
however the divergence problems for zero splitting distance caused by vacuum
fluctuations near a causal or event horizon are genuine and may very well be
the only true divergence problems in LQP.

We have seen that although QM and QFT can be described under a common
mathematical roof ($C^{\ast}$-algebras with a state functional), as soon as
one introduces the physically important localization structure, significant
conceptual differences appear. These differences are due to the presence of
vacuum polarization in QFT as a result of causal localization and they tend to
have dramatic consequences; the most prominent ones will be presented in this
and the subsequent \ sections, as well as in the second part.

The net structure of the observables allows a local comparison of states: two
states are locally equal in a region $\mathcal{O}$ if and only if the
expectation values of all operators in $\mathcal{A(O)}$ are the same in both
states. Local deviations from any state, in particular from the vacuum state,
can be measured in this manner, and states that are indistinguishable from the
vacuum in the causal complement of some region (`strictly localized states'
\cite{Licht}) can be defined. Due to the unavoidable correlations in the
vacuum state in relativistic quantum theory (the Reeh-Schlieder property
\cite{Haag}), the space $H(\mathcal{O})$ obtained by applying the operators in
$\mathcal{A(O)}$ to the vacuum is, for any open region $\mathcal{O}$, dense in
the Hilbert space and thus far from being orthogonal to $H(\mathcal{O}%
^{\prime})$. This somewhat counter-intuitive fact is inseparably linked with a
structural difference between the local algebras and the algebras encountered
in non-relativistic quantum mechanics (or the global algebra of a quantum
field associated with the entire Minkowski space-time) as mentioned in
connection with the breakdown of tensor-factorization (\ref{loc}).

The result is a particular benevolent form of "Murphy's law" for interacting
QFT: \textit{everything which is not forbidden (by superselection rules) to
couple, really does couple}. On the level of interacting particles this has
been termed \textit{nuclear democracy}: Any particle whose superselected
charge is contained in the spectrum resulting from fusing charges in a cluster
of particles can be viewed as a bound state of that cluster. This renders
interacting QFT conceptually much more attractive and fundamental than QM, but
it also contributes to its computational complexity if one tries to access it
using operator or functional methods from QM. The latter method also are
responsible for the occurance of those infinities in the first place which one
then "renormalizes" away by invoking the distinction between Lagrangian and
physical coupling and mass parameters. If one does perturbative QFT according
to its own principles there is never any infinity, but the recursive
implementation of the principles may generate paramters which were not on ones
mind at the beginning (the Epstein-Glaser iteration). 

It is believed that any violation of of the above Murphy's law also violates
the setting of pointlike generated QFT. So the only known approach to particle
physics which is not subject to this law and at least maintains
macro-causality is the before presented quantum mechanical DPI
setting\footnote{It is quite probable that semiinfinite stringlike
interactions violate Murphy's law as well as the crossing property. But even
then they would be still valid in the subalgebra/subspace of local
observables.}. Whereas the latter has minimal projections corresponding to
optimal observations, this is not so for the local algebras which turn out to
be of type III (in the terminology of Murray and von Neumann); in these
algebras every projection is isometrically equivalent to the largest projector
which is the identity operator. Some physical consequences of this difference
have been reviewed in \cite{Yngva}. The claim is not that subalgebras of QM
cannot be of type III but rather that physical subalgebras obtained by the
operator methods of QM (in particular by Born localization) remain of type I.

The Reeh-Schlieder property \cite{Haag} (in more popular but less precise
terminology: the state-field relation) also implies that the expectation value
of a projection operator localized in a bounded region \textit{cannot} be
interpreted as the probability of detecting a particle-like object in that
region, since it is necessarily nonzero if acting on the vacuum state. Our
later study reveals that the restriction of the vacuum (or any other global
finite energy state) to $\mathcal{A(O})$ is entangled in a much more radical
sense than the ground state of a quantum mechanical system under the spatial
inside/outside split. The reduced ground state on $A(\mathcal{O})$ transmutes
into a KMS thermal state at a appropriately normalized (Hawking)
temperature\footnote{The effects we are concerned with are ridiculously small
and probably never mearurable, but here we are interested in principle aspects
of the most successfull and fundamental theory and not in FAPP issues.}. The
intrinsically defined \textit{modular "Hamiltonian"} associated via modular
operator theory to standard pair ($\mathcal{A(O)},\Omega_{vac}$) is always
available in the mathematical sense but allows a physical interpretation only
in those rare cases when it coincides with one of the global spacetime
generators. Well known cases are. the Lorentz boost for the wedge region in
Minkowski spacetime (the Unruh effect) and the positive generator of a
double-cone preserving conformal transformation in a conformal theory. This
phenomenon has the same origin as the later discussed universal area
proportionality of localization entropy which is the entropic side of modular localization).

There exists in fact a whole family \textit{of modular Hamiltonians} since the
operators in $\mathcal{A(O)}$ naturally fulfill the KMS condition of any
standard pair ($\mathcal{A(\check{O})},\Omega_{vac}$) for $\mathcal{\check
{O}\supset O}$: how the different modular thermal states physically "out
themselves" depends on which larger system one wants the operators in
$\mathcal{A(O)}$ to be associated with, i.e. it depends on the
$\mathcal{\check{O}}$-localization of the observers. The original system has
no preference for a particular modular Hamiltonian, it fulfills all those
different KMS properties with respect to all those infinitely many different
modular Hamiltonians $H_{\operatorname{mod}}(\mathcal{O})$ simultaneously. In
certain cases there is a preferred region where this situation of extreme
virtuality caused by vacuum polarization passes to real physics. The most
interesting and prominent case comes about when spacetime curvature is
creating a black hole\footnote{Even in that case there is no difference
whether one associates the localization property with the outside, inside, or
with the horizon of the black hole.}. In such a situation the fleeting "as if"
aspect of a causal localization horizon (e.g. the Unruh horizon) changes to
give room for a more real \textit{event horizon}. For computations of thermal
properties however, including thermal entropy, it does not matter whether the
horizon is a fleeting causal localization horizon or a "real" curvature
generated black hole event horizon. This leads to a picture about the LQP-QG
(quantum gravity) interface which is somewhat different from that in most of
the literature; we will return to these issues in connection with the
presentation of the \textit{split property} in the section on algebraic
modular aspects.

A direct comparison with B-N-W-localization can be made in the case of free
fields which are well defined as operator valued distributions in the space
variables at a fixed time. The one-particle states that are B--N-W localized
in a given space region at a fixed time are not the same as the states
obtained by applying field operators smeared with test functions supported in
this region to the vacuum. The difference lies in the non-local energy factor
$\sqrt{p^{2}+m^{2}}$ linking \ the non-covariant B-N-W states with the states
defined in terms of the covariant field operators.

Causality in relativistic quantum field theory is mathematically expressed
through local commutativity, i.e., mutual commutativity of the algebras
$\mathcal{A(O)}$ and $\mathcal{A(O}^{\prime}$). There is an intimate
connection of this property with the possibility of preparing states that
exhibit no mutual correlations for a given pair of causally disjoint regions.
In fact, in a recent paper Buchholz and Summers \cite{Bu-Su} show that local
commutativity is a necessary condition for the existence of such uncorrelated
states. 

Conversely, in combination with some further properties (split property
\cite{Do-Lo}, existence of scaling limits) that are physically plausible, have
been verified in models and follow from assumption about what constitutes a
physical phase space degree of freedom cardinality in QFT, local commutativity
leads to a very satisfactory picture of statistical independence and local
preparabilty of states in relativistic quantum field theory. We refer to
\cite{Sum}\cite{Wer} for thorough discussions of these matters and
\cite{Yngva}\cite{MSY} for a brief review of some physical consequences. The
last two papers explain how the above mentioned concepts avoids the defects of
the NW localization and resolve spurious problems rooted in assumptions that
are in conflict with basic principles of relativistic quantum physics. In
particular it can be shown how an alleged difficulty \cite{Heger}\cite{Bu-Yn}
with Fermi's famous Gedankenexperiment, which Fermi proposed in order to show
that the velocity of light is also the limiting propagation velocity in
quantum electrodynamics, can be resolved by taking \cite{Yngva} into account
the progress on the conceptual issues of causal localization and the gain in
mathematical rigor since the times of Fermi.

After having discussed some significant conceptual differences between QM and
LQP, one naturally asks for an argument why and in which way QM appears as a
nonrelativistic limit of LQP. The standard kinematical reasoning of the
textbooks is acceptable for fermionic/bosonic systems in the sense of FAPP,
but has not much strength on the conceptual level. To see its weakness,
imagine for a moment that we would live in a 3-dim. world of anyons (abelian
plektons, where plektons are Wigner particles with braid group statistics).
Such relativistic objects are by their very statistics so tightly interwoven
that there simply are no compactly localized free fields which only create a
localized anyon without a vacuum polarization cloud admixture. In such a world
no nonrelativistic limit which maintains the spin-statistic connection could
lead to QM, the limiting theory would rather \textit{remain a nonrelativistic
QFT}. There is simply no Schr\"{o}dinger equation for plektonic particle-like
objects which carry the spin/statistics properties of anyons. In 4-dimensional
spacetime there is no such obstacle against QM, simply because there exist
relativistic free fields whose application to the vacuum generates a
vacuum-polarization-free one-particle state and the spin-statistics structure
does not require the permanence of polarization clouds in the nonrelativistic limit.

\subsection{ Modular localization}

Previously it was mentioned on several occasions that the localization
underlying QFT can be freed from the contingencies of field coordinatizations.
This is achieved by a physically as well mathematically impressive but for
historic and sociological reasons little known theory. Its name "modular
theory" is of mathematical origin and refers to a substantial generalization
of the (uni)modularity encountered in the relation between left/right Haar
measure in group representation theory. In the middle 60s the mathematician
Tomita presented a vast generalization of this theory to operator algebras and
in the subsequent years this theory received essential improvements from Takesaki.

At the same time the physicists Haag, Hugenholtz and Winnink published their
work on statistics mechanics of open systems \cite{Haag}. When the physicists
and mathematicians met at a conference in Baton Rouge in 1966, there was
surprise and satisfaction about the perfection with which these independent
developments supported each other. Physicists not only adapted mathematical
terminology, but mathematicians also took some of their names from physicists
as e.g. KMS states which refer to Kubo, Martin and Schwinger who introduced an
analytic property of Gibbs states merely as a computational tool (in order to
avoid computing traces). Haag Hugenholtz and Winnink realized that this
property (which they termed the KMS property) is the only property which
survives in the thermodynamic limit when the trace formulas becomes divergent.

This turned out to be the right concept for formulating and solving problems
directly in the setting of open systems. In the present work the terminology
is mainly used for thermal states which are not Gibbs density. They are
typical for LQP for example every multiparticle state $\Omega_{particle}$ of
finite energy including the vacuum (i.e. every physical particle state) upon
restriction to a local algebra $\mathcal{A(O})$ becomes a KMS state with
respect to a "modular Hamiltonian" which is canonically determined by
($\mathcal{A(O}),\Omega_{particle}$).

Connes, in his path-breaking work on the classification of von Neumann
factors, made full use of this hybrid math-phys. terminology. Nowadays one can
meet mathematicians who use the KMS property but do not know its origin. One
can hardly think of any other confluence of mathematical and physical ideas on
such a profound and at the same time natural level, even including the
beginnings of QT when the mathematical apparatus needed for QM already existed.

About 10 years later Bisognano and Wichmann \cite{Bi-Wi} discovered that a
vacuum state restricted to a wedge-localized operator algebra $\mathcal{A(}W)$
in QFT defines a modular setting in which the restricted vacuum becomes a
thermal KMS state with respect to the wedge-affiliated L-boost "Hamiltonian".
This step marks the beginning of a very natural yet unexpected relation
between thermal and geometric properties which is totally characteristic for
QFT which is not shared by classical theory nor by QM. Thermal aspects of
black holes were however discovered independent of this work, and the first
physicist who saw the connection was Geoffrey Sewell \cite{Sew}.

The theory becomes more accessible for physicists if one introduces it first
in its more limited spatial- instead of its full algebraic- context. Since as
a foundational structure of LQP it merits more attention than it hitherto
received from the particle physics community, I will present some of its
methods and achievements.

It has been realized by Brunetti, Guido and Longo \cite{BGL} \footnote{With
somewhat different motivations and lesser mathematical rigor also in
\cite{Sch}.} there is a natural localization structure on the Wigner
representation space for any positive energy representation of the proper
Poincar\'{e} group. Upon second quantization this representation theoretical
determined localization theory gives rise to a local net of operator algebras
on the Wigner-Fock space over the Wigner representation space.

The starting point is an irreducible representation $U_{1}$of the Poincar\'{e}%
\'{}%
group on a Hilbert space $H_{1}$ that after "second quantization" becomes the
single-particle subspace of the Hilbert space (Wigner-Fock-space) $H_{WF}$ of
the field\footnote{The construction works for arbitrary positive energy
representations, not only irreducible ones.}. The construction then proceeds
according to the following steps \cite{BGL}\cite{Fa-Sc}\cite{MSY}. To maintain
simplicity we limit our presentation to the bosonic situation.

One first fixes a reference wedge region, e.g. $W_{0}=\{x\in\mathbb{R}%
^{d},x^{d-1}>\left\vert x^{0}\right\vert \}$ and considers the one-parametric
L-boost group (the hyperbolic rotation by $\chi$ in the $x^{d-1}-x^{0}$ plane)
which leaves $W_{0}$ invariant; one also needs the reflection $j_{W_{0}}$
across the edge of the wedge (i.e. along the coordinates $x^{d-1}-x^{0}$). The
Wigner representation is then used to define two commuting wedge-affiliated
operators%
\begin{equation}
\mathfrak{\delta}_{W_{0}}^{it}=\mathfrak{u}(0,\Lambda_{W_{0}}(\chi=-2\pi
t)),~\mathfrak{j}_{W_{0}}=\mathfrak{u}(0,j_{W_{0}})
\end{equation}
where attention should be paid to the fact that in a positive energy
representation any operator which inverts time is necessarily
antilinear\footnote{The wedge reflection $\mathfrak{j}_{W_{0}}$ differs from
the TCP operator only by a $\pi$-rotation around the W$_{0}$ axis.}. A unitary
one- parametric strongly continuous subgroup as $\delta_{W_{0}}^{it}$ can be
written in terms of a selfadjoint generator as $\delta_{W_{0}}^{it}%
=e^{-itK_{W_{0}}}$ and therefore permits an "analytic continuation" in $t$ to
an unbounded densely defined positive operators $\delta_{W_{0}}^{s}$. With the
help of this operator one defines the unbounded antilinear operator which has
the same dense domain.%
\begin{align}
\mathfrak{s}_{W_{0}}  &  =\mathfrak{j}_{W_{0}}\mathfrak{\delta}_{W_{0}}%
^{\frac{1}{2}}\\
\mathfrak{j\delta}^{\frac{1}{2}}\mathfrak{j}\mathfrak{=\delta}  &  ^{-\frac
{1}{2}}%
\end{align}

Whereas the unitary operator $\delta_{W_{0}}^{it}$ commutes with the
reflection, the antiunitarity of the reflection changes causes a change of
sign in the analytic continuation as written in the second line. This leads to
the idempotency of the s-operator on its domain as well as the identity of its
range with its domain
\begin{align*}
\mathfrak{s}_{W_{0}}^{2} &  \subset\mathbf{1}\\
dom~\mathfrak{s} &  =ran~\mathfrak{s}%
\end{align*}
Such operators which are unbounded and yet involutive on their domain are very
unusual; according to my best knowledge they only appear in modular theory and
it is precisely these unusual properties which are capable to encode geometric
localization properties into domain properties of abstract quantum operators.
The more general algebraic context in which Tomita discovered modular theory
will be mentioned later.

The idempotency means that the s-operator has $\pm1$ eigenspaces; since it is
antilinear the +space multiplied with $i$ changes the sign and becomes the -
space; hence it suffices to introduce a notation for just one eigenspace%
\begin{align}
\mathfrak{K}(W_{0})  &  =\{domain~of~\Delta_{W_{0}}^{\frac{1}{2}%
},~\mathfrak{s}_{W_{0}}\psi=\psi\}\\
\mathfrak{j}_{W_{0}}\mathfrak{K}(W_{0})  &  =\mathfrak{K}(W_{0}^{\prime
})=\mathfrak{K}(W_{0})^{\prime},\text{ }duality\nonumber\\
\overline{\mathfrak{K}(W_{0})+i\mathfrak{K}(W_{0})}  &  =H_{1},\text{
}\mathfrak{K}(W_{0})\cap i\mathfrak{K}(W_{0})=0\nonumber
\end{align}

It is important to be aware that, unlike QM, we are here dealing with real
(closed) subspaces $\mathfrak{K}$ of the complex one-particle Wigner
representation space $H_{1}$. An alternative which avoids the use of real
subspaces is to directly deal with complex dense subspaces as in the third
line. Introducing the graph norm of the dense space the complex subspace in
the third line becomes a Hilbert space in its own right. The second and third
line require some explanation. The upper dash on regions denotes the causal
disjoint (which is the opposite wedge) whereas the dash on real subspaces
means the symplectic complement with respect to the symplectic form
$Im(\cdot,\cdot)$ on $H_{1}.$

The two properties in the third line are the defining property of what is
called the \textit{standardness property} of a real
subspace\footnote{According to the Reeh-Schlieder theorem a local algebra
$\mathcal{A(O})$ in QFT is in standard position with respect to the vacuum
i.e. it acts on the vacuum in a cyclic and separating manner. The spatial
standardness, which follows directly from Wigner representation theory, is
just the one-particle projection of the Reeh-Schlieder property.}; any
standard K space permits to define an abstract s-operator%
\begin{align}
\mathfrak{s}(\psi+i\varphi)  &  =\psi-i\varphi\\
\mathfrak{s}  &  =\mathfrak{j}\delta^{\frac{1}{2}}\nonumber
\end{align}
whose polar decomposition (written in the second line) yields two modular
objects, a unitary modular group $\delta^{it}$ and a antiunitary reflection
which generally have however no geometric significance. The domain of the
Tomita s-operator is the same as the domain of $\delta^{\frac{1}{2}}$ namely
the real sum of the K space and its imaginary multiple. Note that this domain
is determined by Wigner group representation theory only.

It is easy to obtain a net of K-spaces by $U(a,\Lambda)$-transforming the
K-space for the distinguished $W_{0}.$ A bit more tricky is the construction
of sharper localized subspaces via intersections
\begin{equation}
\mathfrak{K}(\mathcal{O})=%
{\displaystyle\bigcap\limits_{W\supset\mathcal{O}}}
\mathfrak{K}(W)
\end{equation}
where $\mathcal{O}$ denotes a causally complete smaller region (noncompact
spacelike cone, compact double cone). Intersection may not be standard, in
fact they may be zero in which case the theory allows localization in $W$ (it
always does) but not in $\mathcal{O}.$ Such a theory is still causal but not
local in the sense that its associated free fields are pointlike.

There are three classes of irreducible positive energy representation, the
family of massive representations $(m>0,s)$ with half-integer spin $s$ and the
family of massless representation which consists really of two subfamilies
with quite different properties namely the $(0,h),$ $h$ half-integer class
(often called the neutrino, photon class), and the rather large class of
$(0,\kappa>0)$ infinite helicity representations parametrized by a
continuous-valued Casimir invariant $\kappa$ \cite{MSY}$.$

For the first two classes the $\mathfrak{K}$-space is standard for arbitrarily
small $\mathcal{O}$ but this is definitely not the case for the infinite
helicity family for which the compact localization spaces turn out to be
trivial\footnote{It is quite easy to prove the standardness for spacelike cone
localization (leading to singular stringlike generating fields) just from the
positive energy property which is shared by all three families \cite{BGL}.}.
Their tightest localization, which still permits nontrivial (in fact standard)
$\mathfrak{K}$-spaces for \textit{all} positive energy representations, is
that of a \textit{spacelike cone} with an arbitrary small opening angle whose
core is a \textit{semiinfinite string} \cite{BGL}; after "second quantization
(see next subsection) these strings become the localization region of
string-like localized covariant generating fields\footnote{The epithet
"generating" refers to the tightest localized singular field (operator-valued
distribution) which generates the spacetime-indexed net of algebras in a QFT.
In the case of localization of states the generators are state-valued
distributions.}. The modular localization of states, which is governed by the
unitary representation theory of the Poincar\'{e} group, has only two kind of
generators: pointlike state and semiinfinite stringlike states; generating
states of higher dimensionality (brane states) can be excluded. 

Although the observation that the third Wigner representation class is not
pointlike generated was made many decades ago, the statement that it is
semiinfinite string-generated and that this is the worst possible case of
state localization is of a more recent vintage \cite{BGL} since it needed the
support of the modular theory.

There is a very subtle aspect of modular localization which one encounters in
the second Wigner representation class of massless finite helicity
representations (the photon, graviton..class). Whereas in the massive case all
spinorial fields $\Psi^{(A,\dot{B})}$ the relation of the physical spin $s$
with the two spinorial indices follows the naive angular momentum composition
rules%
\begin{align}
\left\vert A-\dot{B}\right\vert  &  \leq s\leq\left\vert A+\dot{B}\right\vert
,\text{ }m>0\label{line}\\
s  &  =\left\vert A-\dot{B}\right\vert ,~m=0\nonumber
\end{align}
where the second line contains the hugely reduced number of spinorial
descriptions for zero mass and finite helicity although in both cases the
number of pointlike generators which are linear in the Wigner creation and
annihilation operators \cite{MSY}.

By using the recourse of string-localized generators $\Psi^{(A,\dot{B})}(x,e)$
even in those cases where the representation has pointlike generators, one can
even in the massless case return to the full spinorial spectrum as in the
first line (\ref{line}). These generators are covariant and "string-local"%

\begin{align}
U(\Lambda)\Psi^{(A,\dot{B})}(x,e)U(\Lambda) &  =D^{(A,\dot{B})}(\Lambda
^{-1})\Psi^{(A,\dot{B})}(\Lambda x,\Lambda e)\\
\left[  \Psi^{(A,\dot{B})}(x,e),\Psi^{(A^{\prime},\dot{B}^{\prime})}%
(x^{\prime},e^{\prime}\right]  _{\pm} &  =0,~x+\mathbb{R}_{+}e><x^{\prime
}+\mathbb{R}_{+}e^{\prime}\nonumber
\end{align}
Here the unit vector $e$ is the spacelike direction of the semiinfinite string
and the last line expresses the spacelike fermionic/bosonic spacelike
commutation. The best known illustration is the ($m=0,s=1$) representation; in
this case it is well-known that although a generating pointlike field strength
exists, there is no pointlike vectorpotential. The modular localization
approach offers as a substitute a stringlike covariant vector potential
$A_{\mu}(x,e).$ In the case ($m=0,s=2$) the "field strength" is a fourth
degree tensor which has the symmetry properties of the Riemann tensor; in fact
it is often referred to as the linearized Riemann tensor. In this case the
string-localized potential is of the form $g_{\mu\nu}(x,e)$ i.e. resembles the
metric tensor of general relativity. The consequences of this localization for
a reformulation of gauge theory will be mentioned in a separate subsection.

A different kind of spacelike string-localization arises in d=1+2 Wigner
representations with anomalous spin \cite{Mu1}. The amazing power of this
modular localization approach is that it preempts the spin-statistics
connection in the one-particle setting, namely if s is the spin of the
particle (which in d=1+2 may take on any real value) then one finds for the
connection of the symplectic complement with the causal complement the
generalized duality relation
\[
\mathfrak{K}(\mathcal{O}^{\prime})=Z\mathfrak{K}(\mathcal{O})^{\prime}%
\]
where the square of the twist operator $Z=e^{\pi is}~$is easily seen (by the
connection of Wigner representation theory with the two-point function) to
lead to the statistics phase: $Z^{2}=$ statistics phase \cite{Mu1}. The fact
that one never has to go beyond string localization (and fact, apart from
those mentioned cases, never beyond point localization) in order to obtain
generating fields for a QFT is remarkable in view of the many attempts to
introduce extended objects into QFT.

It should be clear that modular localization which goes with real subspaces
(or dense complex subspaces) unlike B-N-W localization cannot be connected
with probabilities and projectors. It is rather related to causal localization
aspects and the standardness of the K-space for a compact region is nothing
else then the one-particle version of the Reeh-Schlieder property. It was
certainly the kind of localization which Wigner was looking for because it
represents the caminho real from representation theory into QFT. As will be
seen in the next section it is also an important tool in the non-perturbative
construction of interacting models.

\subsection{Algebraic aspects of modular theory}

A net of real subspaces $\mathfrak{K}(\mathcal{O})$ $\subset$ $H_{1}$ for an
finite spin (helicity) Wigner representation can be "second
quantized"\footnote{The terminology 2$^{nd}$ quantization is a misdemeanor
since one is dealing with a rigorously defined functor within QT which has
little in common with the artful use of that parallellism to classical theory
called "quantization". In Edward Nelson's words: (first) quantization is a
mystery, but second quantization is a functor.} via the CCR (Weyl)
respectively CAR quantization functor; in this way one obtains a covariant
$\mathcal{O}$-indexed net of von Neumann algebras $\mathcal{A(O)}$ acting on
the bosonic or fermionic Fock space $H=Fock(H_{1})$ built over the
one-particle Wigner space $H_{1}.$ For integer spin/helicity values the
modular localization in Wigner space implies the identification of the
symplectic complement with the geometric complement in the sense of
relativistic causality, i.e. $\mathfrak{K}(\mathcal{O})^{\prime}%
=\mathfrak{K}(\mathcal{O}^{\prime})$ (spatial Haag duality in $H_{1}$). The
Weyl functor takes the spatial version of Haag duality into its algebraic
counterpart. One proceeds as follows: for each Wigner wave function
$\varphi\in H_{1}$ the associated (unitary) Weyl operator is defined as%
\begin{align}
Weyl(\varphi)  &  :=expi\{a^{\ast}(\varphi)+a(\varphi)\},Weyl(\varphi)\in
B(H)\\
\mathcal{A(O})  &  :=alg\{Weyl(\varphi)|\varphi\in\mathfrak{K}(\mathcal{O}%
)\}^{^{\prime\prime}},~~\mathcal{A(O})^{\prime}=\mathcal{A(O}^{\prime
})\nonumber
\end{align}
where $a^{\ast}(\varphi)$ and $a(\varphi)$ are the usual Fock space creation
and annihilation operators of a Wigner particle in the wave function $\varphi
$. We then define the von Neumann algebra corresponding to the localization
region $\mathcal{O}$ in terms of the operator algebra generated by the
functorial image of the modular constructed localized subspace $\mathfrak{K}%
(\mathcal{O})$ as in the second line. By the von Neumann double commutant
theorem, our generated operator algebra is weakly closed by definition.

The functorial relation between real subspaces and von Neumann algebras via
the Weyl functor preserves the causal localization structure and hence the
spatial duality passes to its algebraic counterpart. The functor also commutes
with the improvement of localization through intersections $\cap$ according to
$K(\mathcal{O})=\cap_{W\supset O}K(W),~\mathcal{A(O})=\cap_{W\supset
O}\mathcal{A}(W)$ as expressed in the commuting diagram%
\begin{align}
&  \left\{  K(W)\right\}  _{W}\longrightarrow\left\{  \mathcal{A}(W)\right\}
_{W}\\
&  \ \ \downarrow\cap~~~\ \ \ \ \ \ \ \ \ \ ~\ ~\downarrow\cap\nonumber\\
~~  &  \ \ \ K(\mathcal{O})\ \ \ \longrightarrow\ \ ~\mathcal{A(O})\nonumber
\end{align}
Here the vertical arrows denote the tightening of localization by intersection
whereas the horizontal ones denote the action of the Weyl functor.

The case of half-integer spin representations is analogous \cite{Fa-Sc}, apart
from the fact that there is a mismatch between the causal and symplectic
complements which must be taken care of by a \textit{twist operator}
$\mathcal{Z}$ and as a result one has to use the CAR functor instead of the
Weyl functor.

In case of the large family of irreducible zero mass infinite spin
representations in which the lightlike little group is faithfully represented,
the finitely localized K-spaces are trivial $\mathfrak{K}(\mathcal{O})=\{0\}$
and the \textit{most tightly localized nontrivial spaces} \textit{are of the
form} $\mathfrak{K}(\mathcal{C})$ for $\mathcal{C}$ a \textit{spacelike cone}.
As a double cone contracts to its core which is a point, the core of a double
cone is a \textit{covariant spacelike semiinfinite string}. The above
functorial construction works the same way for the Wigner infinite spin
representation except that there are no nontrivial algebras which have a
smaller localization than $\mathcal{A(C})$ and there are no fields which are
sharper localized than a semiinfinite string. Point- (or string-) like
covariant fields are singular generators of these algebras i.e. they are
operator-valued distributions. Stringlike generators, which are also available
in the pointlike case, turn out to have an improved short distance behavior;
whereas e.g. the short distance dimension of a free pointlike vectorfield is
$sddA_{\mu}(x)=2$ for its stringlike counterpart one has $sddA_{\mu}%
(x,e)=1~$\cite{MSY}. They can be constructed from the unique Wigner
representation by so called intertwiners between the canonical and the many
possible covariant (dotted-undotted spinor finite representations of the
L-group) representations. The Euler-Lagrange aspect plays no role in these
construction since the causal aspect of hyperbolic differential propagation
are fully taken care of by modular localization.

A basis of local covariant field coordinatizations is then defined by Wick
composites of the free fields. The string-like fields do not follow the
classical behavior since already before introducing composites one has a
continuous family of not classically intertwiners between the unique Wigner
infinite spin representation and the continuously many covariant string
interwiners. Their non-classical aspects, in particular the absence of a
Lagrangian, are the reason why their spacetime description in terms of
semiinfinite string fields has been discovered only recently and not at the
time of Jordan's field quantization.

Using the standard notation $\Gamma$ for the second quantization functor which
maps real localized (one-particle) subspaces into localized von Neumann
algebras and extending this functor in a natural way to include the images of
the $\mathfrak{K}(\mathcal{O})$-associated $s,\delta,j$ which are denoted by
$S,\Delta,J$ one arrives at the Tomita Takesaki theory of the interaction-free
local algebra ($\mathcal{A(O}),\Omega$) in standard position\footnote{The
functor $\Gamma$ preserves the standardness i.e. maps the spatial one-particle
standardness into its algebraic counterpart.}%
\begin{align}
&  H_{Fock}=\Gamma(H_{1})=e^{H_{1}},~\left(  e^{h},e^{k}\right)
=e^{(h,k)}\label{mod}\\
&  \Delta=\Gamma(\delta),~J=\Gamma(j),~S=\Gamma(s)\nonumber\\
&  SA\Omega=A^{\ast}\Omega,~A\in\mathcal{A}(O),~S=J\Delta^{\frac{1}{2}%
}\nonumber
\end{align}

With this we arrive at the core statement of the Tomita-Takesaki theorem which
is a statement about the two modular objects $\Delta^{it}$ and $J$ on the
algebra%
\begin{align}
\sigma_{t}(\mathcal{A(O}))  &  \equiv\Delta^{it}\mathcal{A(O})\Delta
^{-it}=\mathcal{A(O})\\
J\mathcal{A(O})J  &  =\mathcal{A(O})^{\prime}=\mathcal{A(O}^{\prime})\nonumber
\end{align}
in words: the reflection $J$ maps an algebra (in standard position) into its
von Neumann commutant and the unitary group $\Delta^{it}$ defines an
one-parametric automorphism-group $\sigma_{t}$ of the algebra. In this form
(but without the last geometric statement involving the geometrical causal
complement $\mathcal{O}^{\prime})$ the theorem hold in complete mathematical
generality for standard pairs ($\mathcal{A},\Omega$). The free fields and
their Wick composites are "coordinatizing" singular generators of this
$\mathcal{O}$-indexed net of algebras in the sense that the smeared fields
$A(f)$ with $suppf\subset\mathcal{O}$ are (unbounded operators) affiliated
with $\mathcal{A(O}).$

In the above second quantization context the origin of the T-T theorem and its
proof is clear: the symplectic disjoint passes via the functorial operation to
the operator algebra commutant and the spatial one-particle automorphism goes
into its algebraic counterpart. The definition of the Tomita involution $S$
through its action on the dense set of states (guarantied by the standardness
of $\mathcal{A}$) as $SA\Omega=A^{\ast}\Omega$ and the action of the two
modular objects $\Delta,J$ (\ref{mod}) is part of the general setting of the
modular Tomita-Takesaki theory; standardness is the mathematical terminology
for the Reeh-Schlieder property i.e. the existence\footnote{In QFT any finite
energy vector (which of course includes the vacuum) has this property as well
as any nondegenerated KMS state. In the mathematical setting it is shown that
standard vectors are "$\delta-$dense" in $H$.} of a vector $\Omega\in H$ with
respect to which the algebra acts cyclic and has no "annihilators" of
$\Omega.$ Naturally the proof of the abstract T-T theorem in the general
setting of operator algebras is more involved.

The important property which renders this useful beyond free fields as a new
constructive tool in the presence of interactions, is that for $\left(
\mathcal{A}(W),\Omega\right)  ~$ the antiunitary involution $J$ depends on the
interaction, whereas $\Delta^{it}$ continues to be uniquely fixed by the
representation of the Poincar\'{e} group i.e. by the particle content. In fact
it has been known for some \cite{Sch} time that $J$ is related with its free
counterpart $J_{0}$ through the scattering matrx%
\begin{equation}
J=J_{0}S_{scat} \label{scat}%
\end{equation}

This modular role of the scattering matrix as a relative modular invariant
between an interacting theory and its free counterpart comes as a surprise. It
is precisely this role which opens the way for an inverse scattering
construction .

The physically relevant facts emerging from modular theory can be compressed
into the following statements\footnote{Alain Connes would like to see a third
spatial decomposition in that list namely the decomposition of $K$ into a
certain positive cone and its opposite. With such a requirement one could
obtain the entire \textit{algebra strucure from that of states}. This
construction has been highly useful in Connes classification of von Neumann
algebras, but it has not been possible to relate this with physical concepts.}

\begin{itemize}
\item \textit{The domain of the unbounded operators }$S(\mathcal{O})$\textit{
is fixed in terms of intersections of the wedge domains associated to
}$S(W);~$\textit{in other words it is determined by the particle content alone
and therefore of a kinematical nature. These dense domains change with
}$\mathcal{O}$\textit{ i.e. the dense set of localized states has a bundle
structure.}

\item \textit{The complex domains }$DomS(\mathcal{O})=K(\mathcal{O}%
)+iK(\mathcal{O})$\textit{ decompose into real subspaces }$K(\mathcal{O}%
)=\overline{\mathcal{A(O})^{sa}\Omega}.$\textit{ This decomposition contains
dynamical information which in case }$\mathcal{O}=W$\textit{ reduces to the
S-matrix (\ref{scat}). Assuming the validity of the crossing properties for
formfactors, the S-matix fixes }$\mathcal{A}(W)$\textit{ uniquely \cite{S3}.}
\end{itemize}

The remainder of this subsection contains some comments about a remarkable
constructive success of these modular methods. For this we need some
additional terminology. Let us enlarge the algebraic setting by admitting
unbounded operators with Wightman domains which are affiliated to
$\mathcal{A(O})$ and just talk about "$\mathcal{O}$-localized operators" when
we do not want to distinguish between bounded and affiliated unbounded
operators. We call an $\mathcal{O}$-localized operators a vacuum
\textbf{p}olarization \textbf{f}ree \textbf{g}enerator (PFG) if applied to the
vacuum it generated a one-particle state without admixture of a
vacuum-polarization cloud. The following two theorems have turned out to be
useful in a constructive approach based on modular theory.

\textbf{Theorem}: \textit{The existence of an }$\mathcal{O}$\textit{-localized
PFG for a causally complete subwedge region }$\mathcal{O}\subset W$\textit{
implies the absence of interactions i.e. the generating fields are (}a slight
generalization of the Jost-Schroer theorem \cite{Wigh} which used the
existence of pointlike covariant fields).

\textbf{Theorem (}\cite{BBS}\textbf{)}: \textit{Modular theory for wedge
algebras insures the existence of PFGs even in the presence of interactions.
Hence the wedge region permits the best compromise between interacting fields
and one-particle states.}

\textbf{Theorem (}\cite{BBS}\textbf{)}: \textit{Wedge localized PFGs with
Wightman-like domain properties ("tempered" PFGs) lead to the absence of
particle creation (pure elasic }$S_{scat})$\textit{ which in turn is only
possible in d=1+1 and leads to the factorizing models (which hitherto were
studied in the setting of the bootstrap-formfactor program). The compact
localized subalgebra }$\mathcal{A(O})$\textit{ have no PFGs and possess the
full interaction-induced vacuum polarization clouds.}

Some additional comments will be helpful. The first theorem gives an intrinsic
(i.e. not dependent on any Lagrangian or other extraneous properties) local
definition of the presence of interaction, although it is \ not capable to
differentiate between different kind of interactions (which would be reflected
in the shapes of interaction-induced polarization clouds). The other two
theorems suggest that the knowledge of the wedge algebra $\mathcal{A(}%
W)\subset B(H)$ may serve as a useful starting point for classifying and
constructing models of LQP in a completely intrinsic fashion\footnote{In
particular the above commuting square remains valid in the presence of
interactions if one changes $\mathcal{O}\rightarrow W.$}. Knowing generating
operators of $\mathcal{A(}W)$ including their transformation properties under
the Poincar\'{e} group is certainly sufficient and constitutes the most
practical way for getting the construction started.

All wedge algebras possess affiliated PFGs but only in case they come with
reasonable domain properties ("tempered") they can presently be used in
computations. This requirement only leaves models in d=1+1 which in addition
must be factorizing (integrable); in fact the modular theory used in
establishing these connections shows that there is a deep connection between
integrability in QFT and vacuum polarization properties \cite{BBS}.

Tempered PFGs which generate wedge algebra for factorizing models have a
rather simple algebraic structure. They are of the form%
\begin{equation}
Z(x)=\int\left(  \tilde{Z}(\theta)e^{-ipx}+h.c.\right)  \frac{dp}{2p_{0}}
\label{Z}%
\end{equation}
where in the simplest case $\tilde{Z}(\theta),\tilde{Z}^{\ast}(\theta)~$are
one-component objects\footnote{This case leads to the Sinh-Gordon theory and
related models.} which obey the Zamolodchikov-Faddeev commutation relations.
In this way the formal Z-F device which encoded the two-particle S-matrix into
the commutation structure of the Z-F algebra receives a profound spacetime
interpretation. Like free fields they are on mass shell, but their creation
and annihilation part obeys the Z-F commutation relations instead of the
standard free field relations; as a result they are incompatible with
pointlike localization but turn out to comply with wedge localization
\cite{S3}.

The simplicity of the wedge generators in factorizing models is in stark
contrast to the richness of compactly localized operators e.g. of operators
affiliated to a spacetime double cone $\mathcal{D}$ which arises as a relative
commutant $\mathcal{A(D})=\mathcal{A}(W_{a})^{\prime}\cap\mathcal{A}(W)$. The
wedge algebra $\mathcal{A}(W)$ has simple generators and the full space of
formal operators affiliated with $\mathcal{A(}W)$ has the form of an infinite
series in the Z-F operators with coefficient functions $a(\theta_{1}%
,...\theta_{n})$ with analyticity properties in a $\theta$-strip%
\begin{equation}
A(x)=%
{\displaystyle\sum}
\frac{1}{n!}%
{\displaystyle\int_{\partial S(0,\pi)}}
d\theta_{1}...%
{\displaystyle\int_{\partial S(0,\pi)}}
d\theta_{n}e^{-ix\sum p(\theta_{i})}a(\theta_{1},...\theta_{n})\tilde
{Z}(\theta_{1})...\tilde{Z}(\theta_{1}) \label{Zam}%
\end{equation}
where for the purpose of a compact notation we view the creation part
$\tilde{Z}^{\ast}(\theta)$ as $\tilde{Z}(\theta+i\pi)$ i.e. as coming from the
upper part of the strip $S(0,\pi)$\footnote{The notation is suggested by the
the strip analyticity coming from wedge localization. Of course only certain
matrix elements and expectation values, but not field operators or their
Fourier transforms, can be analytic; therefore the notation is symbolic.}. The
requirement that the formal expressions of the form commutes with a series of
the same kind but translated by $a$ defines formally a subspace of operators
affiliated with $\mathcal{A(D})=\mathcal{A}(W_{a})^{\prime}\cap\mathcal{A}%
(W).$ As a result of the simplicity of the $\tilde{Z}$ generators one can
characterize these subspaces in terms of analytic properties of the
coefficient functions $a(\theta_{1},...\theta_{n})$ and one recognizes that
they are identical \cite{Sch} to the so-called formfactor postulates in the
bootstrap-formfactor approach \cite{Ba-Ka}.

This is similar to the old Glaser-Lehmann-Zimmermann representation for the
interacting Heisenberg field \cite{GLZ} in terms of incoming free field (in
which case the spacetime dependent coefficient functions turn out to be
on-shell restrictions of Fourier transforms of retarded functions), except
that instead of the on-shell incoming fields one takes the on-shell $\tilde
{Z}$ operators and the coefficient functions are the (connected part of the)
multiparticle formfactors. As was the case with the GLZ series, the
convergence of the formfactor series has remaind an open problem. So unlike
perturbative series resulting from renormalized perturbation theory which have
been shown to diverge even in models with optimal short distance behavior
(even Borel resummability does not help), the status of the GLZ and formfactor
series remains unresolved.

The main property one has to establish if one's aim is to secure the existence
of a QFT with local observables, is the standardness of the double cone
intersection $\mathcal{A(D})=\cap_{W\supset\mathcal{D}}\mathcal{A(}W).$ Based
on nuclearity properties of degrees of freedom in phasespace discovered by
Buchholz and Wichmann \cite{Bu}, Lechner has found a method within the modular
operator setting of factorizing models which achieves precisely this
\cite{Lech1}\cite{Lech2}. For the first time in the history of QFT one now has
a construction method which goes beyond the Hamiltonian- and
measure-theoretical approach of the 60s \cite{G-J}. The old approach could
only deal with superrenormalizable models i.e. models whose basic fields did
not have a short distance dimension beyond that of a free field.

At this point it is instructive to recall that although QFT has been the most
successful of all physical theories as far as observational predictions are
concerned, in comparison to those theories which already have a secure place
in the pantheon of theoretical physics, it remains quite shaky concerning its
mathematical and conceptual foundations. Looking at the present sociological
situation it seems that the last past success which led to the standard model
has generated an amnesia about foundational problems. Post standard model
theories as string theory profited from this situation.

The factorizing models form an interesting battle ground where problems, which
accompanied QFT almost since its birth, have a good chance to receive a new
push. The very existence of these theories, whose fields have anomalous
trans-canonical short distance dimensions with interaction-dependent
strengths, shows that there is nothing intrinsically threatening about
singular short distance behavior. Whereas in renormalized perturbation theory
the power counting rule only permits logarithmic corrections to the canonical
(free field) dimensions, the construction of factorizing models starting from
wedge algebras and their $Z$ generators permits arbitrary high powers. That
many problems of QFT are not intrinsic but rather caused by a particular
method of quantization had already been suspected by the protagonist of QFT
Pascual Jordan who, as far back as 1929, pleaded for a formulation "without
(classic) crutches" \cite{Jo}. The above construction of factorizing models
which does not use any of the quantization schemes and in which the model does
not even come with a Lagrangian name may be considered at the first
realization of Jordan's plea at which he arrived on purely philosophically grounds.

The significant conceptual distance between QM and LQP begs the question in
what sense the statement that QM is a nonrelativistic limit of LQP must be
understood. By this we do not mean a manipulation in a Lagrangian or
functional integral representation, but an argument which starts from the
correlation functions or operator algebras of an interacting LQP. Apparently
such an argument does not yet exist. One attempt in this direction could
consist in starting from the known formfactors of a factorizing model as e.g.
the Sinh-Gordon model and see the simplifications (vanishing of the vacuum
polarization contributions) for small rapidity $\theta.$ An understanding in
this sense would be an essential improvement of our understanding of the
QM-QFT interface.

Since modular theory continues to play an important role in the two remaining
sections, some care is required in avoiding potential misunderstandings. It is
very crucial to be aware of the fact that by restricting the global vacuum
state to, a say double cone algebra $\mathcal{A(D}),$ there is no change in
the values of the global vacuum expectation values%
\begin{equation}
(\Omega_{vac},A\Omega_{vac})=\left(  \Omega_{\operatorname{mod},\beta}%
,A\Omega_{\operatorname{mod},\beta}\right)  ,~\text{ }A\in\mathcal{A(D})
\end{equation}
where for the standard normalization of the modular Hamiltonian\footnote{The
modular Hamiltonian lead to fuzzy motions within $\mathcal{A(O})$ except in
case of $\mathcal{O}=W$ when the modular Hamiltonian is identical to the boost
generator.} $\beta=1.$ This notation on the right hand side means that the
vacuum expectation values, if restricted to $A\in\mathcal{A(D}),$ fulfill an
additional property (which without the restriction to the local algebra would
not hold), namely the KMS relation%
\begin{equation}
\left(  \Omega_{\operatorname{mod},\beta},AB\Omega_{\operatorname{mod},\beta
}\right)  =\left(  \Omega_{\operatorname{mod},\beta},B\Delta_{\mathcal{A(O}%
)}A\Omega_{\operatorname{mod},\beta}\right)  \label{ther}%
\end{equation}
At this point one may wonder how a global vacuum state can turn into a thermal
state on a smaller algebra without any thermal exchange taking place. The
answer is that the in terms of ($A(\mathcal{D}),\Omega_{vac}$) canonically
defined modular Hamiltonian $H_{\operatorname{mod}}$ with $\Delta
=e^{-H_{\operatorname{mod}}}$ is very different from the original translative
Hamiltonian $H_{tr}$ whose lowest energy eigenstate defines the vacuum whereas
$H_{\operatorname{mod}}$ is not the translative heat bath Hamiltonian but
rather a thermal Hamiltonian which describes a (in general "fuzzy") movement
inside $\mathcal{D}$ with the restricted vacuum representing a homogeneous
thermal mean eigenstate. Its vanishing eigenvalue is not that of a ground
state but sits in the middle of a symmetric two-sided spectrum. What has
changed through the process of restriction is not the state but rather the way
of looking at it: $H_{\operatorname{mod}}$ describes the dynamics of an
"observer" confined to $\mathcal{D}$.

The thermal aspect of modular theory does of course not mean that one is
converting a ground state into a thermal state in the sense of creating heat.
As in the case of QM where the subdivision into an inside and outside the
region via Born localization is primarily a Gedanken-act in order to create
information theoretical entanglement from states which have a finite energy
above the vacuum, the thermal aspect of the modular localization serves in
first place to find an efficient description of the vacuum in terms of a
smaller causally closed world. There is nothing more precise and intrinsic
than saying that the restricted vacuum is a KMS state of a certain Hamiltonian
even if there is no physical realization. In both cases one views the original
state but from a different viewpoint.

But what about the Unruh effect which is associated with the Rindler wedge
$\mathcal{O}=W$ $?$ Isn't this more than just a change of viewpoint? Yes and
no. In order to create such a horizon the observer must be uniformely
accelerated which requires feeding energy into the system. In other words the
innocent looking restriction requires an enormous expenditure thus revealing
in one example what is behind the innocent sounding word "restriction". An
accepted rule of thumb is that only when the modular Hamiltonian describes a
movement which corresponds to a diffeomorphism of spacetime is there a chance
to think in terms of an Unruh kind of Gedankenexperiment. The modular
situation is more advantageous in black hole situation where the position of
event horizons is fixed by the metric. For example there exists a pure state
on the Kruskal extension of the Schwarzschild solution (the Hartle-Hawking
state) which restricted to the outside of the black hole describes the
timelike Killing movement; in this case there is no doubt that the restriction
corresponds to the natural time development in the world outside the black hole.

In fact there is a \textit{continuous family of modular "Hamiltonians"} which
are the generators the modular unitaries for sequences of included regions.
The modular Hamiltonian of the larger region will of course spread the smaller
localized algebra into the larger region. All these modular Hamiltonians have
the two-sided symmetric spectrum which is typical for Hamiltonians in a KMS
representation \cite{Haag}.

Besides the thermal description of restricted states there is one other
macroscopic manifestation of vacuum polarization which has caused unbelieving
amazement namely the cyclicity of the vacuum (the Reeh-Schlieder property)
with respect to algebras localized in arbitrarily small spacetime region. The
idea that by doing something in a small earthly laboratory for a say small
fraction of a second by which a state "behind the moon" maybe approximated
with arbitrary precision is certainly such a statement.

Both consequences of vacuum polarization, the thermal aspect of state
restriction and the "state behind the moon property" are manifestations of a
holistic behavior which is unheard of in QM. Instead of the division into an
object to be measured on and the environment without which the modern quantum
mechanical measurement theory can hardly be formulated, one has a situation
which makes such a dichotomy illusory. By restricting to the inside one
already specifies the dynamics on the causal disjoint, it governed by the same
Hamiltonian. In the state behind the moon argument the difficulty in a
system-environmen dichotomyt is even more palpable.

This is indeed an extremely surprising feature which goes considerably beyond
the kinematical change caused by entanglement as the result of the quantum
mechanical division into measured system and environment. It is this
dependence of the reduced vacuum state on the localization region inside which
it is tested with localized algebras which raises doubts about what one really
associates with the non-fleeting persistent properties of a material
substance. The monad description in the next section strengthens this little
holistic aspect of LQP.

In both cases QM as well as QFT the entanglement comes about by a different
ways of looking at the system and not by changing intrinsic properties, but
the \textit{thermal entanglement} of QFT associated with modular localization
is more spectacular than the (\textit{"cold") information-theoretic kind of
entanglement} of QM associated with Born localization.

As we have seen the thermal aspects of modular localization are very rich from
an epistemic viewpoint. The ontic content of these observations is quite weak;
it is only when the (imagined) causal localization horizons passes from a
Gedanken objects to a (real) event horizons through the curvature of spacetime
that the fleeting aspect of observers horizons becomes an ontic property of
spacetime as in black holes. But even if one's main interest is to do black
hole physics, it is wise to avoid a presently popular "shut up and compute"
attitude and to understand the conceptual basis in LQP of the thermal aspect
of localization and the peculiar thermal entanglement which contrasts the
information-theoretical quantum mechanical entanglement. Not caring about
these conceptual aspects one may easily be drawn into a fruitless and
protractive arguments as it happened (and is still happening) with the
entropy/information loss issue. These problems are connected to an
insufficient conceptual understanding of QFT by identifying it with some sort
of relativistic QM.

Up to now the terminology "localization" was used both for states and for
subalgebras. Whereas in the absence of interactions it is true that they are
synonymous in the sense that when a dense subspace of $\mathcal{O}$ localized
states results from the application of an $\mathcal{O}$ localized algebra onto
the vacuum, such a close relation between algebraic and spatial localization
breaks down in the presence of interactions. It is perfectly conceivable to
have a theory with "topological charges" \cite{Haag} which by definition are
not compactly localizable but rather spacelike cone localizable (in terms of
generating fields semiinfinite string-localizable). In that case only the
neutral observable algebra has the usual compact localizability. property
whereas the charge-carrying part of the total algebra is at best localizable
in the sense of semiinfinite strings (the field generators) of such an
algebra. But massive charged states can always be written in terms of
pointlike state-valued distributions; the modular decomposition theory of
representations of the Poincar\'{e} group prevents pointlike generation only
in the presence of infinite spin representations.

\subsection{String-localization and gauge theory}

Zero mass fields of finite helicity play a crucial role in gauge theory.
Whereas in classical gauge theory a pointlike massless vectorpotential is not
unphysical because otherwise it would contradict classical principles but
rather because it is a pure auxiliary construct in the setting of Maxwell's
theory, The situation changes radically in QFT because a covariant zero mass
vectorpotential can only exist in form of a string-localized field, a
covariant pointlike localized contradicts the Hilbert space structure of QT.
Nevertheless there exists an indefinite formalism with additional "ghost
degrees of freedom", the Gupta Bleuler formalism in QED and the BRST ghost
formalism in QCD, which permits to return to physical quantities in a Hilbert
space setting by what is interpreted as the quantum version of gauge invariance.

This only has been shown in perturbation theory and it would not be over
pessimistic to expect that manipulations which depend on convergence have no
meaning outside the Hilbert space topology. But there is actually a much
stronger physical reason having to do with localization why the gauge theory
does not give the full insight into QED and this certainly worsens in passing
to QCD. In QED the physically most important objects are the electrically
charged fields. Since they are nonlocal physical object they are not part of
the perturbative gauge setting which aims at the local gauge invariant; in
some sense they are nonlocal gauge invariant fields but this is just another
way of saying that their construction requires ingenuity and luck because the
formalism of perturbation theory does not lead to a computational rule for
charged fields.

Using a version of perturbation theory which was especially designed for
charged fields, Steinmann \cite{Stein} succeeded to make some headway on this
problem. Related to this nonlocality aspect is the subtle relation of
electrically charged fields to charged particles is their involved infrared
aspect; a charged particle even after a long time of having left the
scattering region will never be without an infinite cloud of infrared real
(not virtual!) photons whose energy is below the registering resolution and
which therefore remain "invisible". This makes charge particles
"infraparticles" i.e. objects whose scattering theory does not lead to
scattering amplitudes but only the inclusive cross sections.

The problem of nonlocal fields becomes much more series in theories involving
vectorfields coupled among themselves. Whereas one believes to have a physical
understanding of the local gauge invariant composites whose perturbation
expansion has incurable infrared divergencies, there is not the slightest idea
what is at the place of the charged fields. For four decades one uses nice
words as quark and gluon confinement well aware that, different from QM, QFT
has no mechanism which can enclose quantum matter in a vault, rather the only
modality to arrive at "invisibility" is through still stronger delocalization,
involving not only the would be charged matter but even the selfinteracting
vectorpotentials. There is little chance that this can be done "by hand", as
in the case of QED.

The only alternative to the present gauge method in which the pointlike
localization of covariant vectorpotentials is paid for by the unphysical ghost
formalism and the subsequent restriction to local observables is to work with
string-localized potentials $A_{\mu}(x,e).$ This poses completely new and
still largely unsolved problems. But before commenting on this new task, it is
helpful to delineate what one expects of such an alternative approach.

Superficially such string-localized fields seem to be indistinguishable from
the axial gauge; here as there the conditions $\partial^{\mu}A_{\mu
}(x)=0=e^{\mu}A_{\mu}(x)$ are obeyed. In the axial gauge interpretation the
$e$ is a gauge parameter and does not participate in Lorentz transformations
whereas in the formula for a string-localized field the spacelike unit vector
transforms as a string direction. The distance between the two concepts
increases when one passes from free fields to their perturbative interactions.
It is well known that the axial gauge formalism fails on its infrared
divergencies; there has been no successful computation involving loops. The
string-localized approach on the other hand requires the computation of
perturbative correlation functions with variable string direction%
\begin{equation}
\left\langle 0\left\vert A_{\mu_{1}}(x_{1},e_{1})...A_{\mu_{n}}(x_{n}%
,e_{n})\psi(y_{1})...\bar{\psi}(y_{m})\right\vert 0\right\rangle
\end{equation}
Also at the end of the calculation when one extracts the physical result one
must study the infrared behavior of coalescing strings, it is important to
keep the directions $e_{i}$ as variable string directions (and not as a fixed
gauge parameter) and consider the correlations as distributions in $x$ as well
as in $e.$ Only in this way one has a chance to handle the infrared
divergencies of these theories and understand their physical role. Note that
the implicit dependence of the matter field $\psi$ and $\bar{\psi}$ on the
$e^{\prime}s$ of the inner vectorpotential lines has been omitted for obvious
reasons of compactness of notation.

The infrared singularity will appear at the end when all $e^{\prime}s$
coalesce but then it is not just an unspecified divergence but rather takes on
the appearance of a short distance limit in a one lower dimensional de Sitter
space (the space of spacelike directions) This spacetime interpretation for
the infrared divergencies is missing in the axial gauge setting.

Although the power counting behavior of quadrilinear interactions of
string-localized fields is not worse than that of pointlike interactions in
the ghost formalism of gauge theory, there is a serious problem with the
perturbative Epstein Glaser \cite{E-G} iteration. In the pointlike case the
knowledge of the n$^{th}$ order determines the n+1 order up to a term on the
total diagonal which limits the freedom to the addition of pointlike
composites. The presence of string-like fields invalidates this argument; even
if all strings lie in one spacelike hypersurface, the freedom is larger than
that allowed by the total diagonal. What one hopes for is that the freedom can
be described in terms of some string composites, whatever that means.

For QED one is in the fortunate position of being able to compare whatever
this new string-localized perturbation leads to with the before mentioned
gauge theoretic calculation where the nonlocal aspect of electrically charged
operators needs to be added by hand. In the QCD case it would be unreasonable
to expect that nonlocal physical operators do not exist, but nobody in the 4
decades since the inception of gauge QFT came up with a reasonable suggestion
as to how such object could look like. This problem has been studied on the
lattice, but if lattice gauge theory was not even able to describe the
charge-carrying infraparticle fields of QED what hope is left to understand
the structure of QCD beyond its pointlike-localized observable content? It
seems that the only possibility to make headway on or most important problems
of particle physics is to go after a theory of interacting semiinfinite
string-localized fields.

Although for massive particles there is no structural reason to introduce
string-localized fields, viz. the free massive vectorpotential $A_{\mu}(x)$
which exists in the Wigner-Fock space, the short distance dimension of such
fields increase with increasing spin so that already a s=1 field leads to
sdd=2 and hence does not allow a fourth power coupling of fields within the
power counting limit. A string-localized massive vectorpotential has sdd=1 and
would therefore lead to quadrilinear interactions within this limit. Assuming
that the Epstein-Glaser iteration has an extension to string-localized fields
one could hope for a better intrinsic understanding of the Schwinger Higgs
screening mechanism than within the present gauge setting. In particular one
may finally understand how the presence of a scalar particle renders the whole
theory pointlike local so that the use of a string-localized vector potential
had the sole purpose of enabling a renormalizable interaction.

\subsection{Building LQP via \textit{positioning of monads} in a Hilbert
space}

We have seen that modular localization of states and algebras is an intrinsic
i.e. field-coordinatization-independent way to formulate the kind of
localization which is characteristic for QFT. It is deeply satisfying that it
also leads to a new constructive view of QFT.

\textbf{Definition }(Wiesbrock \cite{Wies1})\textbf{:} \textit{An inclusion of
standard operator algebras }$\left(  \mathcal{A\subset B},\Omega\right)
$\textit{ is "modular" if }$\left(  \mathcal{A},\Omega\right)  $\textit{ and
(}$\mathcal{B}$\textit{,}$\Omega$\textit{) are standard and }$\Delta
_{\mathcal{B}}^{it}$\textit{ acts like a compression on }$\mathcal{A}$\textit{
i.e. }$Ad\Delta_{\mathcal{B}}^{it}\mathcal{A}\subset\mathcal{A}.$\textit{ A
modular inclusion is said to be standard if in addition the relative commutant
(}$\mathcal{A}^{\prime}\cap\mathcal{B},\Omega$\textit{) is standard. If this
holds for }$t<0$\textit{ one speaks about a -modular inclusion.}

The study of inclusions of operator algebras has been an area of considerable
mathematical interest. Particle physics uses 3 different kind of inclusions;
besides the modular inclusions which play the principal role in this section
there are \textit{split inclusions} and inclusions with conditional
expectations or using the name of their creator Vaughn \textit{Jones
inclusions.} Split inclusions play an important role in structural
investigation and are indispensable in the study of thermal aspects of
localization notably localization entropy (see next chapter). Jones inclusions
result from reformulating the DHR theory of superselection sectors which in
its original formulation uses the formalism of localized endomorphisms of
observable algebras.

The important achievement of that theory is that the local system of
observables has enough structure in order to complement the theory with its
charged fields and their inner symmetries such that the original observables
reemerge as the fixed point under this symmetry. This projection is
accomplished in terms of a conditional expectation. The prototype of a
conditional expectation in the conventional formulation of QFT (based on the
use of charge-carrying fields) is the averaging over the compact internal
symmetry group with its normalized Haar measure ($U(g)$ denotes the
representation of the internal symmetry group)
\begin{align}
\mathcal{A}  &  =\int d\mu(g)AdU(g)\mathcal{F}\\
E  &  :\mathcal{F}\overset{\mu}{\longrightarrow}\mathcal{A},~E^{2}=E\nonumber
\end{align}
i.e. the conditional expectation $E$ projects the (charged) field algebra
$\mathcal{F}$ onto the (neutral) observable algebra $\mathcal{A}$ and such
inclusions which do not change the localization are therefore related to
internal symmetries as opposed to spacetime symmetries.

Inclusions $\mathcal{A}\subset\mathcal{B}$ with conditional expectation
$E(\mathcal{B})$ cannot be modular and the precise understanding why this is
the case discloses interesting insights. According to a theorem of Takesaki
\cite{Tak} the existence of a conditional expectation is tantamount to the
modular group of the smaller algebra being equal to the restriction of that of
the bigger. Hence the natural generalization of this situation is that the
group $Ad\Delta_{\mathcal{B}}^{it}$ of the larger algebra acts on
$\mathcal{A}$ for either $t<0$ or for $t>0$ as a compression (endomorphism)
and the absence of a conditional expectation. Intuitively speaking modular
inclusions are too deep in order to allow conditional expectations. Continuing
this line of speculative reasoning one would expect that as "flat" inclusions
with conditional expectations are related to inner symmetries, "deep"
inclusions of the modular kind lead to spacetime symmetries.

Surprisingly this rough guess turns out to be amazingly correct. The main aim
of modular inclusions is really to \textit{generate spacetime symmetry} as
well as the \textit{net of spacetime indexed algebras} which are covariant
under these symmetries. This is done as follows: from the two modular groups
$\Delta_{\mathcal{B}}^{it},\Delta_{\mathcal{A}}^{it}$ one can form a unitary
group $U(a)$ which together with the modular unitary group of the smaller
algebra $\Delta_{\mathcal{B}}^{it}$ leads to\ the commutation relation
$\Delta_{\mathcal{B}}^{it}U(a)=U(e^{-2\pi t}a)\Delta_{\mathcal{B}}^{it}$ which
characterizes the 2-parametric translation-dilation (Anosov) group. One also
obtains a system of local algebras by applying these symmetries\ to the
relative commutant $\mathcal{A}^{\prime}\cap\mathcal{B}.$ From these relative
commutants one may form a new algebra $\mathcal{C}$%
\begin{equation}
\mathcal{C}\equiv\overline{\bigcup_{t}Ad\Delta_{\mathcal{B}}^{it}%
(\mathcal{A}^{\prime}\cap\mathcal{B})}%
\end{equation}
In general $\mathcal{C}\subset\mathcal{B}$ and we are in a situation of a
nontrivial inclusion to which the Takesaki theorem is applicable (the modular
group of $\mathcal{C}$ is the restriction of that of $\mathcal{B})$ which
leads to a conditional expectation $E:\mathcal{B}\rightarrow\mathcal{C}$;
$\mathcal{C}$ may\ also be trivial. The most interesting situation arises if
the modular inclusion is \textit{standard }i.e. all three algebras
$\mathcal{A},\mathcal{B},\mathcal{A}^{\prime}\cap\mathcal{B}$ are standard
with respect to $\Omega;$ in that case we arrive at a chiral QFT.

\textbf{Theorem}: (Guido,Longo and Wiesbrock \cite{G-L-W}) \textit{Standard
modular inclusions are in one-to-one correspondence with strongly additive
chiral LQP.}

Here chiral LQP is a net of local algebras indexed by the intervals on a line
with a Moebius-invariant vacuum vector and \textit{strongly additive} refers
to the fact that the removal of a point from an interval does not
\textquotedblleft damage\textquotedblright\ the algebra i.e. the von Neumann
algebra generated by the two pieces is still the original algebra. One can
show via a dualization process that there is a unique association of a chiral
net on $S^{1}=\mathbb{\dot{R}}$ to a strongly additive net on $\mathbb{R}$.
Although in our definition of modular inclusion we have not said anything
about the nature of the von Neumann algebras, it turns out that the very
requirement of the inclusion being modular forces both algebras to be
hyperfinite type III$_{1}$ factor algebras. The closeness to Leibniz's idea
about (physical) reality of originating from relations between monads (with
each monad in isolation of being void of individual attributes) more than
justifies our choice of name; besides that "monad" is much shorter than the
somewhat long winded mathematical terminology "hyperfinite type III$_{1}$
Murray-von Neumann factor algebra". The nice aspect of chiral models is that
one can pass between the operator algebra formulation and the construction
with pointlike fields without having to make additional technical
assumptions\footnote{The group theoretic arguments which go into that theorem
\cite{Joerss} seem to be available for any conformal QFT.}. Another
interesting constructive aspect is that the operator-algebraic setting permits
to establish the existence of algebraic nets in the sense of LQP for all $c<1$
representations of the energy-momentum tensor algebra. This is much more than
the vertex algebra approach is able to do since that formal power series
approach is blind against the dense domains which change with the localization regions.

The idea of placing the monad into modular positions within a common Hilbert
space may be generalized to more than two copies. For this purpose it is
convenient to define the concept of a \textit{modular intersection} in terms
of modular inclusion.

\textbf{Definition (}Wiesbrock \cite{Wies1}\textbf{)}: \textit{Consider two
monads }$A$\textit{ and }$B$\textit{ positioned in such a way that their
intersection }$A\cap B$\textit{ together with A and B are in standard position
with respect to the vector }$\Omega\in H$\textit{. Assume furthermore}%

\begin{align}
&  (\mathcal{A}\cap\mathcal{B\subset}\mathcal{A)~}and~(\mathcal{A}%
\cap\mathcal{B\subset B)~}are~\pm mi\\
&  J_{\mathcal{A}}\lim_{t\rightarrow\mp}\Delta_{\mathcal{A}}^{it}%
\Delta_{\mathcal{B}}^{-it}J_{\mathcal{A}}=\lim_{t\rightarrow\mp}%
\Delta_{\mathcal{B}}^{it}\Delta_{\mathcal{A}}^{-it}\nonumber
\end{align}
\textit{then (}$A,B,\Omega$\textit{) is said to have the }$\pm$\textit{
modular intersection property (}$\pm~$\textit{mi)}.

It can be shown that this property is stable under taking commutants i.e. if
$\left(  \mathcal{A},\mathcal{B},\Omega\right)  \pm mi$ then $\left(
\mathcal{A}^{\prime},\mathcal{B}^{\prime},\Omega\right)  $ is $\mp mi.$

The minimal number of monads needed to characterize a 2+1 dimensional QFT
through their modular positioning in a joint Hilbert space is three. The
relevant theorem is as follows

\textbf{Theorem}: (Wiesbrock \cite{Wies2}) \textit{Let }$A_{12}$%
\textit{,}$A_{13}$\textit{ and }$A_{23}$\textit{ be three monads\footnote{As
in the case of a modular inclusion, the monad property is a consequence of the
modular setting. But for the presentation it is more convenient and elegant to
talk about monads from the start.} which have the standardness property with
respect to }$\Omega\in H$\textit{. Assume furthermore that}%
\begin{align}
&  (\mathcal{A}_{12},\mathcal{A}_{13},\Omega)~is~-mi\\
&  (\mathcal{A}_{23},\mathcal{A}_{13},\Omega)~is~+mi\nonumber\\
&  (\mathcal{A}_{23},\mathcal{A}_{12}^{\prime},\Omega)~is~-mi\nonumber
\end{align}

\textit{then the modular groups }$\Delta_{12}^{it}$\textit{, }$\Delta
_{13}^{it}$\textit{ and }$\Delta_{23}^{it}$\textit{ generate the Lorentz group
}$SO(2,1)$\textit{.}

Extending this setting by placing an additional monad $\mathcal{B}$ into a
suitable position with respect to the $\mathcal{A}_{ik}$ of the theorem, one
arrives at the Poincar\'{e} group $\mathcal{P}(2,1)$ \cite{Wies3}$.$ The
action of this Poincar\'{e} group on the four monads generates a spacetime
indexed net i.e. a LQP model and all LQP have a monad presentation.

To arrive at d=3+1 LQP one needs 6 monads. The number of monads increases with
the spacetime dimensions. Whereas in low spacetime dimensions the algebraic
positioning is natural within the logic of modular inclusions, in higher
dimensions it is presently necessary to take some additional guidance from
geometry, since the number of possible modular arrangements for more than 3
monads increases.

We have presented these mathematical results and used a terminology in such a
way that the relation to Leibniz philosophical view is highly visible.

Since this is not the place to give a comprehensive account but only to direct
the attention of the reader to this (in my view) startling conceptual
development in the heart of QFT.

Besides the radically different conceptual-philosophical outlook on what
constitutes QFT, the modular setting offers new methods of construction. It
turns out that for that purpose it is more convenient to start from one monad
$\mathcal{A}\subset B(H)$ and assume that one knows the action of the
Poincar\'{e} group via unitaries $U(a,\Lambda)$ on $\mathcal{A}.$ If one
interprets the monad $\mathcal{A}$ as a wedge algebra $\mathcal{A=}$ than the
Poincar\'{e} action generates a net of wedge algebras $\left\{  \mathcal{A(}%
W\mathcal{)}\right\}  _{W\in\mathcal{W}}.$ A QFT is supposed to have local
observables and if the double cone intersections\footnote{Double cones are the
typical causally complete compact regions which can be obtained by
intersecting wedges.} $\mathcal{A(}D\mathcal{)}$ turn out to be trivial
(multiples of the identity algebra) the net of wedge algebras does not leads
to a QFT. This is comparable to the non-existence of a QFT which was to be
associated via quantization to a Lagrangian. If however these intersections
are nontrivial than the ontological status is much better than that we would
have an existence proof which is much more than a non-converging renormalized
perturbative series of which we do not know if and how it is related to a QFT.
There are of course two obvious sticking points: (1) to find
Poincar\'{e}-covariant generators of \ $\mathcal{A(}W_{0}\mathcal{)}$\ and (2)
a method which establishes the non-triviality of intersections of wedge
algebras and leads to formulas for their generating elements.

As was explained in the previous section, both problems have been solved
within a class of factorizing models. Nothing is known about how to address
these two points in the more general setting i.e. when the tempered PFG are
not available. Perhaps one should first test a perturbative version of this
program which is expected to incorporate more possibilities than the
perturbation theory based on pointlike fields since wedge-localized generators
are free of those ultraviolet aspects which come from pointlike localization.
The dynamic input in that case would not be a Lagrangian but rather the lowest
order (tree-approximation) S-matrix interpreted as the in-out formfactor of
the identity operator.

\subsection{The split inclusion}

There is one property of LQP which is indispensable for understanding how the
quantum mechanical tensor factorization can be reconciled with modular
localization: the \textit{split property}.

\textbf{Definition: }\textit{Two monads }$A,B$\textit{ are in a split position
if the inclusion of monads }$\mathcal{A}\subset\mathcal{B}^{\prime}$\textit{
admits an intermediate type I factor }$\mathcal{N}$\textit{ such that
}$A\subset N\subset B^{\prime}$

Split inclusions are very different from modular inclusions or inclusions with
conditional expectations (Jones-DHR). The main property of a split inclusion
is the existence an $\mathcal{N}$-dependent unitarily implemented isomorphism
of the $\mathcal{A},\mathcal{B}$ generated operator algebra into the tensor
product algebra%
\begin{equation}
\mathcal{A}\vee\mathcal{B\rightarrow A}\otimes\mathcal{B\subset N}%
\otimes\mathcal{N}^{\prime}=B(H) \label{split}%
\end{equation}
The prerequisite for this factorization in the LQP context is that the monads
commute, but it is well-known that local commutativity is not sufficient, the
counterexample being two double cones which touch each other at a spacelike
boundary \cite{Haag}. As soon as one localization region is separated from the
other by a (arbitrary small) spacelike security distance, the interaction-free
net satisfies the split property under very general conditions. In \cite{BAF}
the relevant physical property was identified in form of a phase space
property. Unlike QM, the number of degrees of freedom in a finite phase space
volume in QFT is not finite, but its infinity is quite mild; it is a nuclear
set for free theories and this nuclearity requirement\footnote{A set of
vectors is nuclear if it is contained in the range of a trace class operator.}
is then postulated for interacting theories. The physical reason behind this
nuclearity requirement is that it allows to show the existence of temperature
states once one knows that a QFT exists in the vacuum representation.

The split property for two securely causally separated algebras has a nice
physical interpretation. Let $\mathcal{A}=\mathcal{A(O}),~\mathcal{B}^{\prime
}=\mathcal{A(\check{O}}),~\mathcal{O\subset\check{O}}.$ Since $\mathcal{N}$
contains $\mathcal{A}$ and is contained in $\mathcal{B}^{\prime}$ (but without
carrying the assignment of a localization between $\mathcal{O}$ and
$\mathcal{\check{O})}$, one may imagine $\mathcal{N}$ as an algebra which
shares the sharp localization with $\mathcal{A(O})$ in $\mathcal{O}$, but its
localization in the "collar" between $\mathcal{O}$ and $\mathcal{\check{O}}$
is "fuzzy" i.e. the collar subalgebra is like a "fog" which does not really
occupy the collar region. This is precisely the region which is conceded to
the vacuum polarization cloud in order to spread and thus avoid the infinite
compression into the surface of a sharply localized monad. If we take a
sequence of $\mathcal{N}$'s which approach the monad $\mathcal{A}$ the vacuum
polarization clouds become infinitely large so that no direct definition of
e.g. their energy or entropy is possible.

The inclusion of the tensor algebra of monads into a type I tensor product
(\ref{split}) looks at first sight like a d\'{e}j\`{a} vu of QM tensor
factorization, but there are interesting and important differences. In QM the
tensor factorization obtained from the Born localization projector and its
complement is automatic since the vacuum of QM (or the ground state of a
quantum mechanical zero temperature finite density system) tensor factorizes.
In QFT the vacuum does not tensor factorize at all but there are other "split
vacuum" states in the Hilbert space which emulate a vacuum in the sense that
expectation values of operators in $\mathcal{A(O})\vee\mathcal{A(\check{O}%
}^{\prime})$ factorize in the split vacuum%
\begin{equation}
\left\langle 0_{split}\left\vert AB\right\vert 0_{split}\right\rangle
=\left\langle 0\left\vert A\right\vert 0\right\rangle \left\langle 0\left\vert
B\right\vert 0\right\rangle ,~A\in\mathcal{A(O}),B\in\mathcal{A(\check{O}%
}^{\prime})
\end{equation}
But there is a huge conceptual difference to the quantum mechanical Born
factorization of the "nothing" state. The splitting process requires the
supply of energy since the split vacuum has infinite vacuum polarization (with
finite mean energy) in the collar region which is spacelike to $\mathcal{O}%
\vee\mathcal{\check{O}}^{\prime}.$ If one agrees that the physical states of
QFT are the states with a finite particle number than a split vacuum which has
finite mean energy but infinite particle number does not look very realistic.

The problem of physical realizability is not given much attention in
foundational discussions of QM. But in QFT this issue is more serious since
the situations are much more counter-intuitive.as was shown before with the
state behind the moon argument for the global vacuum. This property is lost in
a split vacuum state but it is unclear how such states can be prepared and
monitored. .

Most foundational properties of QM as violation of Bell's inequalities, the
Schroedinger cat property and many other strong deviations from classical
reality can be experimentally verified. This is not possible for the vacuum
polarization caused properties which result from modular localization since
macroscopic manifestations are too small.. A typical example is the Unruh
effect i.e. the thermal manifestation of a uniformly accelerated particle
counter in the global vacuum where the temperature created by an acceleration
of 1m/sec is $10^{-19}K$ too small for ever being registered. But for the
perception of the reality which underlies LQP the difficulty in registering
such effects does not diminish their importance.

The characterization of the restriction of the global vacuum to a local
algebra in terms of a thermal state for a modular Hamiltonian holds
independent of whether the local algebra is a sharply localized monad
$\mathcal{A(O)}$ or a type I factor $\mathcal{N}$ as above in the splitting
construction. The only difference is that the in the second case the KMS state
is also a Gibbs state i.e. the Hamiltonian has a discrete spectrum (in case
$\mathcal{\check{O}}$ is compact). This thermal reinterpretation of reduced
states does not only hold for the vacuum but applies to all states which are
of physical relevance in particle physics i.e. to all finite energy states for
which the Reeh-Schlieder theorem applies.

Since KMS states on type I factors are Gibbs states, there exists a density
matrix. Therefore these Gibbs state can have a finite energy and entropy
content which for monads is impossible. But a monad may be approximated by a
sequence of type I factors in complete analogy to the thermodynamic limit. In
fact the thermodynamic limit is the only place where a monad algebra appears
in a QM setting; an indication that this limit is accompanied by a qualitative
change is the fact that one looses the density matrix nature of the Gibbs
state which changes to a more singular KMS state which simply does not exist
on quantum mechanical type I algebras. A related fact is the breakdown of the
tensor factorization into physical degrees of freedom and their "shadow world"
which is the basis of the "Thermofield formalism", monad algebras simply do
not allow such a tensor factorization.

The structural difference can be traced back to the modular Hamiltonians,
whereas for monads the modular Hamiltonian has continuous spectrum (a typical
example is a quantum mechanical Gibbs state box Hamiltonian in the
V$\rightarrow\infty$ thermodynamic limit representation) and hence an
ill-defined (infinite) value of energy and entropy, this is not the case for
the $\mathcal{N}$-associated density matrix constructed from the split
situation. So the way out is obvious: just imitate the thermodynamic limit by
constructing a sequence of type I factors (a "funnel") $N_{i}\supset
\mathcal{A(O})$ (by tightening the split) which converge from the outside
towards the monad (equivalently one may approximate from the inside). This is
precisely what will be done for the computation of the localization entropy in
the next section.

making the split limit in which a monad (the limiting KMS equilibrium
situation in the standard heat bath setting) is approximated by a sequence of
finite volume Gibbs states for which energy and entropy are finite and only
diverge in the "monad limit". Indeed this will be the main idea or the
derivation of the entropical area law in the next section.

In the above form the monad-positioning aims at characterizing LQP in
Minkowski spacetime. This begs the question whether there is a generalization
to curved spacetime. A very special exploratory attempt in this direction
would be to investigate whether the Diff(S$^{1}$) symmetries beyond the
Moebius group in chiral theories have a modular origin in terms of positioning
monads relative to reference states. Since the extended chiral theories which
originate from null-surface holography (and not from chiral projections of a
two-dimensional conformal QFT) seem to have great constructive potential, this
question may also be of practical interest. There are indications that this
can be done if one relaxes on the idea of a universal vacuum reference state
and allows "partial vacua" i.e. modular defined states which have geometric
properties only on certain subalgebras (work in progress).

I expect that by pursuing the algebraization of QFT in CST via the positioning
of monads to its limits one will learn important lessons about the true QFT/QG
interface. A conservative approach which explores unknown aspects of QFT while
staying firmly rooted in known principles seems to be the most promising path
in the present situation.

\section{Problematization of the QFT-QG interface}

In the previous section we outlined a radical new way of interpreting the
conceptual content of QFT by highlighting those structure which are most
different from QM. However in doing this we paid attention that this new way
is at the same time conservative vis-\`{a}-vis the underlying physical
principles. In certain cases, as 2-dim. integrable models, where one finds
sufficiently well behaved generators of wedge algebras with simple vacuum
polarization properties, one arrives at a nonperturbative scheme for the
construction of models. The interesting aspect of these constructions, besides
the fact that they are the first existence proofs for strictly renormalizable
models\footnote{The models constructed in the 60s were superrenormalizable
i.e. of a short distance type which does not occur in d=4.}, is that the
umbilical quantization cord with classical physics has been cut, i.e. for the
first time the more fundamental QFT was constructed without any reference to a
quantization parallelism (Lagrangians, Functional Integrals,..).

We also indicated how the positioning of monads could be useful for a better
future understanding of the interface between QFT in CST and QG. In this
section more light will be shed on the thermal manifestations of causal
localization. In particular two recent results about presently hotly debated
topics will be presented namely the \textit{universal area law of localization
entropy,} which shifts\footnote{This applies only to people who thought that
one needs QG in order to understand the area law of black hole entropy.} the
interface between LQP and the elusive QG, and an intrinsic definition of the
\textit{energy density in cosmological reference states} (vacuum-like states
in cosmological models) in the setting of QFT in CST.

\subsection{Some history of area behavior of localization induced vacuum
polarization}

The phenomenon of vacuum polarization has been the point of departure of many
metaphors of which the steaming broil is perhaps the best known because it
occasionally even entered textbooks. In order to support this image its was
claimed that a short time violation of the energy conservation is supported by
the uncertainty relation. A less metaphoric view comes from locally "banging"
on the vacuum i.e. applying a compactly localized observable to it. \ Such a
banged state is characterized by its n-particle matrix elements for all n and
these n-particle vacuum polarization components are in turn special boundary
values of an analytic n-particle master function whose different out-in
particle distributions obtained from the vacuum polarization component by
crossing are the formfactors of $A.$
\begin{align*}
&  A\left\vert 0\right\rangle \simeq\{\left\langle p_{1},...p_{n}\right\vert
A\left\vert 0\right\rangle \}_{n}\\
&  \overset{cros\sin g}{\rightarrow}\{\left\langle -p_{1},...-p_{k}\right\vert
A\left\vert p_{k+1},...p_{n}\right\rangle \}_{n}%
\end{align*}
where the negative mass shell momenta -p denotes the analytic continuation
which is part of the crossing process. The crossing property follows from the
fact that the extended wedge algebra
\[
\mathcal{A}_{ext}(W)\equiv\mathcal{A}_{out}(W)\vee\mathcal{A}(W)\vee
\mathcal{A}_{in}(W)
\]
is a subalgebra which shares the boost Hamiltonian and that the restriction of
the vacuum to $\mathcal{A}_{ext}(W)$ is a boost-KMS state. In order to
generate a local bang on the vacuum one indeed creates a soup of particles and
although the expectation of the energy in such a bang state is finite a bang
with sharp localization has no limitation on high particle momenta in the
formfactors of a localized operator. With other words none of the formfactors
of such an operator vanishes in any region of momenta of multiparticle space
with the only restriction coming from charge superselection rules.

Whereas in QM, relativistic or not, one has great liberty in manipulating
interactions so that almost any outcome can be accommodated, this is not the
case in QFT. This tightness even show up in theorems about the S-matrix as Aks
theorem: in a 4-dimensional QFT nontrivial elastic scattering is not possible
without the presence of inelastic components \cite{BBS}. For the formfactors
the previous statement in a more popular jargon permits a stronger and more
general formulation in terms of a benevolent Murphy's law: all couplings of
local operators to other channels (in the case of formfactors multiparticle
channels) which are not forbidden by superselection rules actually do occur.
Of course one needs to bang onto the vacuum, there is no "boiling soup" in a
an inertial frame without heating the vacuum stove. The formfactor aspect of a
local operator is perhaps the best QFT illustration of Murphy's law to
particle physics.

Of course vacuum polarization as a concomitant phenomenon of QFT was
discovered a long time before the role of locality it became clear. It is
interesting to reformulate Heisenberg's observation in a lightly more modern
context by defining partial charges by limiting the charged region with the
help of smooth test function. In Heisenberg's more formal setting the partial
charge of a free conserved current in a spatial volume V is defined as
\begin{align}
Q_{V}  &  =\int_{V}j_{0}(x,t)d^{3}x\label{Hei}\\
j_{\mu}(x,t)  &  =:\phi^{\ast}(x,t)\overset{\leftrightarrow}{\partial}_{\mu
}\phi(x,t):\nonumber
\end{align}
Introducing a momentum space cutoff, the norm of $Q_{V}\left\vert
0\right\rangle $ turns out to diverge quadratically which together with the
dimensionlessness of Q is tied to the area proportionality. Hence already on
the basis of a crude dimensional reasoning one finds an area proportionality
of vacuum polarization. The cutoff was the prize to pay for ignoring the
singular nature of the current which is really not an operator but rather an
operator-valued distribution.

The modern remedy is to take care of the divergence by treating the singular
current as an operator-valued distribution. Such calculations have been done
in the 60s by using spacetime test functions which regularize the delta
function at coalescing times and are equal to one inside the ball with radius
$R$ and fall off to zero smoothly between $R$ and $R$+$\Delta R.$ Using the
conservation law of the current one can then show that the action of the
regularized partial charge on the vacuum is compressed to the shell
($R,R+\Delta R)$ and diverges quadratically with $\Delta R\rightarrow0$ i.e.
As expected, the vacuum fluctuations vanish weakly as $R\rightarrow\infty$
(even strongly by enlarging the time smearing support together with $R$) i.e.
the limit converges independent of the special test function
weakly\footnote{Although the norm diverges, the inner product of
$Q_{R}\left\vert 0\right\rangle $ with localized states converges to zero in
compliance with the zero charge of the vacuum.} to the global charge operator%
\begin{equation}
\lim_{R\rightarrow\infty}\int f_{R}(\vec{x})g(t)j_{0}(x,t)d^{4}x=Q
\end{equation}

The problem of localization-entropy is conceptually more involved since
entropy is inherently nonlocal in the sense that it cannot be obtained by a
integrating a pointlike conserved current or any other operator but rather
encodes a holistic aspect of an entire algebra. Nevertheless there is an
algebraic analog of the above test function smearing: the splitting property
\cite{Haag}.

Entropy in QM is an information theoretical concept which measures the degree
of entanglement. The standard situation is bipartite spatial subdivision of a
global system so that global pure states decompose into tensor product states
and superpositions of product states called entangled. The entropy is than a
number computed according in the well-known manner with the von Neumann
prescription from the reduced impure state which results in the standard way
from averaging over the opposite component.

The traditional quantum mechanical way to compute entanglement entropy was
applied to QFT of a halfspace (a Rindler wedge in spacetime) for a system of
free fields in a influential 1984 paper \cite{BKLS}. The starting point was
the assumption that the total Hilbert space factorizes in that belonging to
the halfspace QFT and its opposite. The calculation is ultraviolet divergent
and after introducing a momentum space cutoff $\kappa$ the authors showed that
the cutoff dependence is consistent with an area behavior.%

\begin{equation}
S/A=C\kappa^{2} \label{cut}%
\end{equation}
where in the conformal case $C$ is a constant and $\kappa$ is a momentum space
cutoff and $S/A$ denotes the surface density of entropy. The method of
computation is again the integration over the degrees of freedom of the
complement region and the extraction of the entropy from the resulting reduced
density matrix state whose degree of impurity encodes the measure of the
inside/outside entanglement.

It is easy to see this calculation in analogy to Heisenberg's calculation of
charge polarization (\ref{Hei}). In both cases the starting formula is morally
correct but factually wrong. Neither is the partial charge inside a region
defined by a volume integral nor do, as we know from previous sections, global
states in QFT permit an inside/outside factorization. These incorrect
assumptions create the divergencies which are then kept under the lid by QFT
emergency aid: momentum space cutoff. In both cases dimensional arguments lead
to an area proportionality. But the area appears only as a dimension-saving
factor, there is no direct information that in both cases this behavior comes
from vacuum polarization in a shell near the boundary and there is also no
hint as to what is the correct formulation of the starting assumption. Whereas
in the Heisenberg case the correct definition of the partial charge requires
the test function formalism of pointlike currents, the algebraic counterpart
in the case of entropy is the split property. With our preparation of this
important concept in previous sections we now can plunge into medias res.

\subsection{A modern point of view of localization entropy}

Let us first apply the split idea to a two-dimensional conformal QFT in which
case the double cone is a two-dimensional spacetime region consisting of the
forward and backward causal shadow of a line of length $L$ at $t=0$ sitting
inside region obtained by augmenting the baseline on both sides by $\Delta L.$
As a result of the assumed conformal invariance of the theory, the canonical
split algebra inherits the covariances and hence the entropy of the canonical
split algebra can only be a function of the cross ratio of the 4 points
characterizing the split inclusion%
\begin{align}
S  &  =-tr\rho ln\rho=f(\frac{\left(  d-a\right)  \left(  c-b\right)
}{\left(  b-a\right)  \left(  d-c\right)  })\\
with~a  &  <b<c<d=-L-\Delta L<-L<L<L+\Delta L\nonumber
\end{align}
where for conceptual clarity we wrote the formula for generic position of 4
points. Our main interest is to determine the leading behavior of $f$ in the
limit $\Delta L\rightarrow0$ which is the analog of the thermodynamic limit
$V\rightarrow\infty$ for heat bath thermal systems.

The asymptotic estimate for $\Delta L\rightarrow0$ can be carried out with an
algebraic version of the \textit{replica trick} which uses the cyclic orbifold
construction in \cite{Lo-Fe}. First we write the entropy in the form
\begin{equation}
S=-\frac{d}{dn}tr\rho^{n}|_{n=1},~\rho\in M_{can}\subset\mathcal{A}(L+\Delta
L)
\end{equation}
Then one uses again the split property, this time to map the n-fold tensor
product of $\mathcal{A}(L+\Delta L)$ into the algebra of the line
(conveniently done in the compact $S^{1}$) with the help of the $n^{th}$ root
function $\sqrt[n]{z}.$ The part which is invariant under the cyclic
permutation of the n tensor factors defines the algebraic version \cite{Lo-Fe}
of the replica trick. The transformation properties under Moebius group are
now given in terms of the following subgroup of DiffS$^{1}$ written formally
as%
\begin{align}
&  \sqrt[n]{\frac{\alpha z^{n}+\beta}{\bar{\beta}z^{n}+\bar{\alpha}}},~L_{\pm
n}^{\prime}=\frac{1}{n}L_{\pm n},~L_{0}^{\prime}=L_{0}+\frac{n^{2}-1}{24n}c\\
&  \dim_{\min}=\frac{n^{2}-1}{24n}c\nonumber
\end{align}
where the first line is the natural embedding of the n-fold covering of Moeb
in diffS$^{1}$and the corresponding formula for the generators in terms of the
Virasoro generators. As a consequence the minimal $L_{0}^{\prime}$\ value
(spin, anomalous dimension) is the one in the second line. With this
additional information coming from representation theory we are able to
determine at least the singular behavior of $f$ for coalescing points
$b\rightarrow a,$ $d\rightarrow c$%
\begin{equation}
S_{sing}=-lim_{n\rightarrow1}\frac{d}{dn}\left[  \frac{(d-a)(c-b)}%
{(b-a)(d-c)}\right]  ^{\frac{n^{2}-1}{24n}}=\frac{c}{12}ln\frac{(d-a)(c-b)}%
{(b-a)(d-c)}%
\end{equation}
Since the function is only defined at integer n, one needs to invoke Carlson's theorem.

The resulting entropy formula reads%
\begin{equation}
S_{sing}=\frac{c}{12}\ln\frac{(d-a)(c-b)}{(b-a)(d-c)}=\frac{c}{12}%
ln\frac{L(L+\Delta L)}{\left(  \Lambda L\right)  ^{2}}%
\end{equation}
where $c$ in typical cases is the Virasoro constant (which appears also in the
chiral holographic lightray projection).

This result was previously \cite{S2} obtained by the "inverse Unruh effect"
for chiral theories which is a theorem stating that for a conformal QFT on a
line the KMS state obtained by restricting the vacuum to the algebra of an
interval is unitarily equivalent to a global heat bath temperature state at a
certain geometry-dependent value of the temperature. The chiral inverse Unruh
effect involves a change of length parametrization; the length proportionality
of the heat bath entropy (the well known volume factor) is transformed into a
logarithmic length measure.

Although the inverse Unruh effect is restricted to chiral theories, the
analogy of the heat bath entropy with the localization entropy continues to
exert itself. Below it will be shown that the localization entropy in the
n-dimensional case diverges for $\Delta R\rightarrow0,$ with $\Delta R$ the
splitting distance, as%
\begin{align}
&  E\overset{\Delta R\rightarrow0}{\simeq}\frac{R^{n-2}}{\left(  \Delta
R\right)  ^{n-2}}ln\frac{R^{2}}{\left(  \Delta R\right)  ^{2}}\\
&  V\simeq\left(  \Delta R\right)  ^{n-2}ln\left(  \Delta R\right)  ^{-2}%
\end{align}
The reader will notice the close analogy to the heat bath entropy: the
logarithm corresponds to the lightlike length factor of a lightlike slice of
thickness $\Delta R$ and the inverse power is the analog of a transverse
volume factor in the transformation from the thermodynamic limit to the funnel
limit $\Delta R\rightarrow0$ where the second line expresses the
correspondence between the heat bath volume factor and the divergence factors
of the funnel limit of localization entropy.

Compared with the chiral models which can be controlled quite elegantly with
the replica method, the question of higher dimensional localization entropy
looks more involved. A closer look shows that the problem is not to identify
the relevant density matrix leading to the localization entropy, but rather to
explicitely compute its entropy and come up with a formula which replaces
(\ref{cut}). The localization entropy associated with the double cone geometry
may serve as the most typical illustration. To obtain a finite entropy one
needs a sheet of finite thickness as a vault for the vacuum polarization this
time in an algebraic form rather than test function smearing. For this purpose
one uses the previously presented split property of two monads namely a
smaller double cone algebra of size $R$ inside a bigger (say symmetric around
the origin)
\begin{align}
&  \mathcal{A(D}(R))\subset\mathcal{N}\subset\mathcal{A(D}(R+\Delta R))\\
&  \mathcal{A}(ring)\equiv\mathcal{A(D}(R))^{\prime}\cap\mathcal{A(D}(R+\Delta
R)),~\nonumber\\
&  \mathcal{N=A(D}(R))\vee J_{ring}\mathcal{A(D}(R))J_{ring}\nonumber
\end{align}
where $\mathcal{N}$ is the canonically associated type I algebra in terms of
which there is tensor factorization as in (\ref{split}). The relative
commutant in the second line is of special interest since geometrically it
describes the finite shell region (or rather its causal completion) in which
we expect the vacuum polarization to be localized in that ring. The
restriction of the vacuum to $\mathcal{N}$ is a density matrix state
$\rho_{split}$ and the split entropy is the von Neumann entropy of this mixed
state (there is a corresponding density matrix on $\mathcal{N}^{\prime}$ which
leads to the same entropy).

The only place where the split vacuum deviates significantly from the original
vacuum is on observables in the ring region. This is the origin of the area
proportionality (apart from a logarithmic correction)

The resulting formula is most clear in the conformal case because besides the
length $R$ which determines the hyperface "area" $R^{n-2}$ the only other
dimension carrying parameter is $\Delta R$ so that the entropy is%
\begin{equation}
E=C(n)\frac{R^{n-2}}{\left(  \Delta R\right)  ^{n-2}}\frac{c}{12}%
ln\frac{R(R+\Delta R)}{\left(  \Delta R\right)  ^{2}},~C(0)=1
\end{equation}

The physics of the case of the higher dimensional double cone entropy is
similar since the leading contributions is given by the conformal limit. Note
that the modular temperature is always fixed and generally different from the
physical temperature. For the Unruh effect associated with the boost of a
wedge region W the acceleration of the observer (which belongs to a whole
family of observers) enters $T=2\pi\frac{1}{a}.$ In general the physical
temperature differs from the modular by the "surface gravity".

The reason why we have preferred the double cone instead of the wedge region
whose (modular group is geometric even in the massive case) which is in many
aspects simpler is that the inclusion of two wedge algebras is not split. The
explanation is however very simple, the horizon is not finite since the
transverse area is infinite and hence the would be density matrix resulting
from the inclusion diverges.

\subsection{Remarks on holography}

The special role of null-surfaces as causal boundaries, which define places
around which vacuum polarization clouds form, suggests that there may be more
to expect if one only could make QFT on light-front a conceptually and
mathematically valid concept. That this can be indeed achieved is the result
of holography. Holography clarifies most of the problem which were raised by
its predecessor, the "lightcone quantization" and explains why this method
failed. One of the reasons has to do with short distance behavior since the
naive restriction of fields to space- or light-like submanifolds require the
validity of the canonical quantization formalism i.e. a short distance
dimension not worse than sdd=1.

However the causal localization principle at least in its algebraic
formulation permits to attach to each region the algebra of its causal shadow.
For null-surfaces the situation is better. In that case the observable
algebras indexed by regions on the lightfront are really field-generated and
the field generators are transversely extended chiral fields $C(x,\mathbf{x})$
where $x$ denotes the lightlike coordinate on the lightfront and $\mathbf{x}$
parametrizes the n-2 dimensional transverse submanifold. Their commutation
relations are of the form%
\begin{equation}
\left[  C_{i}(x_{1},\mathbf{x}_{1}),C_{j}(x_{2},\mathbf{x}_{2})\right]
=\delta(\mathbf{x}_{1}-\mathbf{x}_{2})\sum_{k=0}^{m}\delta^{(k)}(x_{1}%
-x_{2})C_{k}(x_{1},\mathbf{x}_{1}) \label{commut}%
\end{equation}
where the number m of operator contributions on the right depends on the scale
dimensions of the two operators on the left hand side. As for standard chiral
fields the scale dimensions are unlimited (no restriction to canonicity as for
equal time commutations)\footnote{There can be higher derivatives in the
transverse direction but they are always even whereas the light-like delta
functions are odd.}. The most useful and characteristic property of lightfront
QFT is the total absence of transverse vacuum polarization, this is how the
area behavior manifests itself in the lightfront generating fields.

The modular localization theory plays a crucial role in the construction of a
local net on the lightfront and its generating fields and for this reason one
must start with operator algebras which is in a standard position with respect
to the vacuum. Since the full lightfront algebra is identical to the global
algebra on Minkowski spacetime one must start with a subregion on the
lightfront and the largest such region is half the lightfront whose causal
completion is the wedge so that it can be seen as the wedge%
\'{}%
s causal (upper) horizon $H(W)$\footnote{This is the quantum version of causal
propagation with characteristic data on $H(W).$ A smaller region on LF does
not cast a causal shadow.}%
\begin{equation}
\mathcal{A}(W)=\mathcal{A}(H(W))
\end{equation}
In the spirit explained in previous sections one the constructs the local
structure of $\mathcal{A}(H(W))$ one intersects and recombines the $W$
algebras which have their horizons on the same lightfront. In 4-dimensional
Minkowski spacetime they are connected by a 7-parametric subgroup of the
10-parametric Poincar\'{e} group containing: 5 transformations which leave W
invariant (the boost, 1 lightlike translation, 2 transverse translations, 1
transverse rotation) and 2 which change W (the two "translations" in Wigner's
Little Group). This is precisely the invariance group of the lightfront. It is
not difficult to see that this net of observable algebras on the lightfront
factorizes in the transverse direction. If this net has pointlike generators
they are necessarily of the kind of transverse extended chiral fields
(\ref{commut}).

For free fields the construction can be done explicitly. Since it is quite
interesting and sheds some light on why the holography works whereas the
lightcone quantization did not succeed the remainder of this section will
present the free field holography.

The crucial property which permits a direct holographic projection is the mass
shell representation of a free scalar field%
\begin{equation}
A(x)=\frac{1}{\left(  2\pi\right)  ^{\frac{3}{2}}}\int(e^{ipx}a^{\ast}%
(p)\frac{d^{3}p}{2p_{0}}+h.c.)
\end{equation}
Using this representation one can directly pass to the lightfront by using
lightfront adapted coordinates \ $x_{\pm}=x^{0}\pm x^{3},~\mathbf{x},$ in
which the lightfront limit $x_{-}=0$ can be taken without causing a divergence
in the p-integration. Using a p-parametrization in terms of the wedge-related
hyperbolic angle $\theta:p_{\pm}=p^{0}+p^{3}\simeq e^{\mp\theta},~\mathbf{p}$
the $x_{-}=0$ restriction of $A(x)$%

\begin{align}
&  A_{LF}(x_{+},\mathbf{x})\simeq\int\left(  e^{i(p_{-}(\theta)x_{+}%
+i\mathbf{px}}a^{\ast}(\theta,\mathbf{p})d\theta d\mathbf{p}+h.c.\right)
\label{LF}\\
&  \left\langle \partial_{x_{+}}A_{LF}(x_{+},\mathbf{x})\partial_{x\prime_{+}%
}A_{LF}(x_{+}^{\prime},\mathbf{x}^{\prime})\right\rangle \simeq\frac
{1}{\left(  x_{+}-x_{+}^{\prime}+i\varepsilon\right)  ^{2}}\cdot
\delta(\mathbf{x}-\mathbf{x}^{\prime})\nonumber\\
&  \left[  \partial_{x_{+}}A_{LF}(x_{+},\mathbf{x}),\partial_{x\prime_{+}%
}A_{LF}(x_{+}^{\prime},\mathbf{x}^{\prime})\right]  \simeq\delta^{\prime
}(x_{+}-x_{+}^{\prime})\delta(\mathbf{x}-\mathbf{x}^{\prime})\nonumber
\end{align}
The justification for this formal manipulation consists in using the fact that
the equivalence class of test function which have the same restriction
$\tilde{f}|_{H_{m}}$ to the mass hyperboloid of mass $m$ is mapped to a unique
test function $f_{LF}$ on the lightfront \cite{Dries}\cite{S1}. It only takes
the margin of a newspaper to verify the identity $A(f)=A(\left\{  f\right\}
)=A_{LF}(f_{LF}).$ But note also that this identity does not mean that the
\thinspace$A_{LF}$ generator can be used in the bulk since the inversion
involves an equivalence class and does not distinguish an individual test
function in the bulk; in fact a localized test function $f(x_{+},\mathbf{x})$
is spread out in the bulk.

This corresponds to the classical causal shadow behavior of characteristic
data on the light front: the causal shadow cast from half the lightfront is
the associated wedge but the restriction to transverse or lightlike compact
data does not improve the bulk localization i.e. the sub $H(W)$ localization
does not improve the bulk localization, it only causes fuzziness. So algebraic
holography from a wedge in the bulk is not invertible. the local substructure
of a wedge algebra $\mathcal{A}(W)$ cannot be fully encoded into
$\mathcal{A}(H(W)$), although the two global algebras are identical. This also
applies to event horizons in curved spacetime and is incompatible with the
idea that the information contained in the local bulk substructure of a region
can be encoded into its horizon (for more remarks see the conclusions).

For the case at hand namely the bulk- and lightfront- generators this
projective nature of holography asserts itself in the fact one cannot
reconstruct from the space of $H(W)$ localized smearing functions the local
substructure of the space of $W$-bulk localized test functions. The projection
can be upgraded to an isomorphism by injecting additional knowledge e.g.
knowledge about how the Poincar\'{e} transformation which are not part of the
7-parametric group act on the lightfront generators. This is a trivial step if
the generators of the holographic projection in case their Poincar\'{e}
covariance is known as in the above case of the $a(p),a^{\ast}(p)~$%
annihilation/creation operators; one only has to apply the $x_{-}$ translation
in order to reconstitute the original bulk generators. It can be shown that
under certain reasonable assumptions of a rather general nature the full bulk
structure can be recovered from knowing the local net of holographic
lightfront projections in different lightfront positions related by
Poincar\'{e} transformations.

Historically the "lightcone quantization" which preceded lightfront holography
shares with the latter part of the motivation namely the idea that by using
lightlike directions one can simplify certain aspects of an interacting QFT.
But as the terminology "quantization" reveals this was unfortunately mixed up
with the erroneous idea that in order to achieve this one needs a new
quantization instead of a radical spacetime reordering of a given abstract
algebraic operator substrate whose Hilbert space is always maintained. As
often such views about QFT results from an insufficient appreciation of the
autonomy of the causal locality principle by not separating it sufficiently
from the contingency of pointlike fields.

Formally mass shell representations also exist for interacting fields. In fact
they appeared shortly after the formulation of LSZ scattering theory and they
were introduced in a paper by Glaser, Lehmann and Zimmermann and became known
under their short name of "GLZ representations". They express the interacting
Heisenberg field as a power series in incoming (outgoing) free fields. In case
there is only one type of particles one has:
\begin{align}
&  A(x)=%
{\displaystyle\sum}
\frac{1}{n!}%
{\displaystyle\idotsint\limits_{V_{m}}}
a(p_{1},...p_{n})e^{i\sum p_{k}x}:A_{in}(p_{1})...A_{in}(p_{n}):\frac
{d^{3}p_{1}}{2p_{10}}...\frac{d^{3}p_{1}}{2p_{10}}\label{GLZ}\\
&  A_{in}(p)=a_{in}^{\ast}(p)~on~V_{m}^{+}~and~a_{in}(p)~on~V_{m}%
^{-}\nonumber\\
&  a(p_{1},...p_{n})_{p_{i}\in V_{m}^{+}}=\left\langle \Omega\left\vert
A(0)\right\vert p_{1},...p_{n}\right\rangle
\end{align}
where the integration extends over the forward and backward mass shell
$V_{m}^{\pm}\subset V_{m}$ and the product is Wick ordered. The coefficient
functions for all momenta on the forward mass shell $V_{m}^{+}$ are the vacuum
polarization components of $A$ and the various formfactors (matrix elements
between in ket and out bra states) of are believed (the crossing property) to
be mass shell boundary values of Fourier-transformed retarded functions.

The convergence status of these series is unknown\footnote{In contrast to the
perturbative expansion which is known to diverge even in the Borel sense, the
convergence status of GLZ had not been settled.}, but it is evident that the
formal lightfront restriction for each term in (\ref{GLZ}) does not cause any
short distance divergence. It is also clear that it is not possible to define
a lightfront restriction on vacuum expectations (Wightman functions), one
really needs to reconstruct the operators and verify the prerequisites for a
mass shell representations as (\ref{GLZ}). In contrast to the algebraic
setting the holography based on the GLZ formula is inherently nonlocal since
it requires the full insight into the nonlocal relation between interacting
and incoming fields. There is however no restriction on the short-distance
dimensions of the fields as there was in the old "lightcone quantization.

The holography of individual fields in the mass shell representation
highlights some interesting problems which are important for autonomous
nonperturbative constructions of models in QFT of the kind i.e. constructions
which do not depend on Lagrangian quantization as those presented after
(\ref{Z}) . The more rigorous algebraic method by its very nature (using
relative commutants) only leads to bosonic holographic projections. This means
that the extended chiral structure on the lightfront only contains integral
values in its short distance spectrum; i.e. the generating fields are of the
kind of the chiral components of two-dimensional conserved currents and
energy-momentum tensors. Hence only a small subalgebra of the bulk
algebra\footnote{Apart from conserved currents whose charges must be
dimensionless, fields are not protected against carrying non-integer short
distance scale dimensions.} associated with transverse extended currents,
energy momentum tensor etc. will appear; there would be no anomalous dimension
field in the algebraic holographic projection.

The obvious conjecture is that the objects belonging to the anomalous
dimensional spectrum which could not pass through the "algebraic holographic
projection filter" can be reconstructed via representation theory of
(extended) chiral observable algebras, a version of the DHR superselection
theory which is particularly well developed in chiral models. The pointlike
field holography based on the mass shell representation supports this idea
that the anomalous dimensions of bosonic bulk fields become holographically
encoded into the spin-statistics and scale dimensions of plektonic (anyonic in
the abelian case) chiral fields. Again the projection carries a lot of
information about the bulk but holographic data on one horizon alone do not
allow a unique inversion i.e. holography on null-surfaces does not lead to an
isomorphism. If on the other hand one would know a GLZ-like representations of
the generating lightfront fields one can obtain the GLZ representations of the
bulk fields simply by a $x_{-}$ translation.

Clearly many of these ideas, as important for the future development of QFT as
they may appear, are not yet mature in the sense of mathematical physics.
Therefore it is good to know that there exists an excellent theoretical
laboratory to test these ideas in a better controlled mathematical setting::
two-dimensional factorizing models and their this time bona fide (no
transverse extension) chiral holographic projection. From a previous section
on modular theory we know that these models have rather simple on-shell wedge
generators $Z(x)$ which still maintain a lot of similarity with free fields.
In that case Zamolodchikov proposed a consistency argument which led to
interesting constructive conjectures about relations between factorizing
models and their critical universality classes represented in form of their
conformal short distance limits.

Conceptual-wise the critical conformal limit is very different from its
holographic projection, the former is a different theory whose Hilbert space
has to be reconstructed from the massless correlation function whereas the
latter is just a reprocessing of spacetime ordering of the original quantum
substrate in the original Hilbert space. Assuming that one knows the chiral
fields on the lightray as a power series in term of the
Z-operators\footnote{From the point of view of chiral models such a
representation is of course somewhat unusual.} one has a unique inversion,
i.e. the holographic projection becomes an isomorphism.

Calculations on two models \cite{Ba-Ka}, the Ising field and the Sinh-Gordon
field, have shown that the universality class method and the holographic
projection lead to identical results\footnote{The consistency of the
holographic lightray projection with the critical limit for factorizing models
was checked in an oral discussion with Michael Karowski..}. Whereas the
anomalous dimension of the sinh-Gordon field can not be computed approximately
in terms of doing the integrals in the lowest terms in the mass shell
contributions, the series for the Ising order field can be summed exactly and
yields the expected number 1/16. This is highly suggestive for reinterpreting
the Zamolodchikov way of relating factorizing models with chiral models as
part of holographic projection which may be used as a rigorous relation
between quantum matter in the bulk and its spacetime re-ordered presentation
on the lightray horizon.

The gain in modular symmetry is perhaps the most intriguing aspect of
holography. In general the modular theory for subwedge localization of bulk
lead to algebraic modular groups which cannot be encoded into diffeomeophisms
of the underlying spacetime manifold; the generators of these groups are at
best pseudo-differential operators. However there are strong indications that
their restriction to the horizon are geometric. This situation is particularly
interesting in generic spacetime manifolds which have no bulk symmetry. \ 

This reinterpretation emphasizes the role which holography is expected to play
in the future development of QFT: introduce a different viewpoint about QFT
which permits to partition the difficult task to construct interacting models
into many less difficult tasks. Certainly chiral models are simpler than any
other model, in fact the classification in terms of families and their
explicit construction has already progressed \cite{Kawa}.

Since our presentations of localization entropy and lightfront holography was
in the setting of Minkowski spacetime where there are only causal horizons but
no event horizons the question arises whether there is any reason to expect
any change on black hole event horizons. In view of speculations about black
hole physics in the literature a more specific question would be is it
conceivable that the holography onto the black hole event horizon becomes an
isomorphism instead of a projection so that the whole world above the horizon
becomes imaged onto the horizon? This sounds a bit like science fiction, after
all the Kruskal extension of the Schwarzschild black hole is of the same
bifurcated kind as the wedge situation. There is really no support for such an
idea from QFT on event horizons or from speculations about QG unless one books
the lack of knowledge as an asset for a speculative idea.

The impossibility to store all information in the bulk into a horizon can
already be seen from what is known about the classical characteristic value
problem i.e. the Cauchy problem on null-surfaces as the lightfront. Whereas
from the local data on the lightfront it is possible to reconstruct the data
on certain semiinfinite regions in the bulk as lightlike slabs with either
infinite extension in lightlike- or spacelike transverse direction, it is not
possible to do this for compactly extended bulk data. This has its precise
analog in the quantum case where the local substructure on the lightfront can
only retrieve the operator algebras indexed by the mentioned semiinfinite
regions. In order to recover the full net of spacetime indexed subalgebras one
either needs to know the action of those Poincar\'{e} symmetries beyond the
7-parametric symmetry subgroup of the lightfront or (in case of CST without
symmetries) the data on more than only one null-surface. An holographic
isomorphism in which the lower dimensional manifold has to carry the burden of
more than the cardinality of degrees of freedom as it would be natural for
that lower spacetime dimension only happens in case of the AdS-CFT
correspondence \cite{Du-Re}.

\subsection{Vacuum fluctuations and the cosmological constant problem}

If there is any calculation which holds the record for predicting a quantity
which comes out way off the astrophysically observed mark, namely by at least
40 orders of magnitudes, it is the estimate for the cosmological constant
based on a quantum mechanical argument of filling particle levels above the
vacuum in a similar spirit as occupying levels up to the Fermi surface for
obtaining the ground state for many body systems at finite density and zero
temperature\footnote{The estimate consists in in filling free energy levels
above the free vacuum state up to a certain cutoff mass $\kappa$ which should
be larger than all the physical masses and smaller than the mass corresponding
to the Planck length. The result of such a calculation leads to an energy
density $\rho_{E}$ $\sim\kappa^{4}.$}. As pointed out by Hollands and Wald
\cite{Ho-Wa}\ such global occupation arguments for computing a local density
contradict the holistic aspect of global reference states in a theory which
fulfills the global covariance principle. the vacuum So the estimate which led
to this a gigantic mismatch between quantum mechanics of free relativistic
particle and the astrophysical reality has no credibility. A "cosmological
constant problem" in the sense of mismatch between particle theory and
cosmological observation does not exist and arguments which have been designed
to find a way out of this problem as e.g. the invocation of an
\textit{anthropic principle} share this lack of credibility..

It is instructive to look first at the problem of the cosmological energy
density i.e. the expectation value of the zero-zero component of the
energy-stress tensor in a state $\varphi$ in a QFT in CST. The standard
argument by which one defines the stress-energy tensor as a composite of a
field is well known for free fields, one starts from the bilocal split-point
expression and takes the coalescing point limit after subtracting the vacuum
expectation value so that the result agrees with the Wick-ordered product. The
resulting stress-energy tensor has all the required properties. Its
expectation value is well-defined on a dense set of states which includes the
finite energy states. But contrary to its classical counterpart, there is a
(unexpected at the time of its discovery \cite{EGJ}) problem with its
boundedness from below since one can find state vectors on which the energy
density $T_{00}(x)$ takes on arbitrarily large negative values.

This had of course led to worries since classical the positivity inequalities
were known to be crucial for questions of stability. It started a flurry of
investigations \cite{Ford} which led to state-independent lower bounds for
fixed test functions $T_{00}(f)$ as well as inequalities on subspaces of test
functions. These inequalities which involve the free stress-energy tensor were
then generalized to curved space time\footnote{For recent publication with
many references see \cite{Few}.}. In the presence of curvature the main
problem is that the definition of $T_{\mu\nu}(x)$ is not obvious since in a
generic spacetime there is no vacuum like state which is distinguished by its
high symmetry; and to play that split point game with an arbitrarily chosen
state will not produce a locally covariant energy stress tensor.

A strategy to do this was given in 1994 by Wald \cite{Wa 1994} in the setting
of free fields. His postulates gave rise to what is nowadays referred to as
the \textit{local covariance principle} which is a very nontrivial
implementation of Einstein's classical covariance principle of GR to quantum
matter in curved spacetime (after freeing the classical principle from its
physically empty coordinate invariance interpretation). determines the correct
energy-momentum tensor up to local curvature terms (whose degree depends on
the spin of the free fields). To formulate it one needs to consider all
Lorentz mannifolds with a certain causality structure simultaneously
\cite{BFV}.

In fact one can construct a basis of composite fields so that every member is
a locally covariant composite of the free field such that for the Minkowski
spacetime we re-obtain the simpler Wick basis. The formulation of the local
covariance principle uses local isometric diffeomorphisms of the kind which
already appeared in Einstein's classical formulation and this requires to
consider simultaneously all QFT which share the same quantum substrate but
follow different spacetime ordering principles. In other words, even if one's
interest is to study QFT in a particular spacetime (Robertson-Walker for the
rest of this section), one is forced to look at all globally hyperbolic
spacetimes in order to find the most restrictive condition imposed by the
local covariance principle.

The result is somewhat surprising in that this principle cannot be implemented
by taking the coincidence limit after subtracting the expectation in one of
the states of the theory. Rather one needs to subtract a "Hadamard parametrix"
\cite{Wald2} i.e. a function which depends on a pair of coordinates and is
defined in geometric terms; in the limit of coalescence it depends only on the
metric in a neighborhood of the point of coalescence. Only then the global
dependence on the metric carried by states can be eliminated in favor of a
local covariant dependence on $g_{\mu\nu}(x)$ and its derivatives. As a result
the so-constructed stress-energy tensor at the point $x$ depends only on the
metric in an infinitesimal neighborhood of $x$.\newline.

Recently these renormalization ideas were applied to computations of
backreactions of a scalar massive free quantum field in a spatially flat
Robertson-Walker model. As a substitute for a vacuum state one uses a state of
the Hadamard form since these states fulfill a the so-called microlocal
spectrum condition which emulates the spectrum condition in Minkowski
spacetime. The singular part of a Hadamard state is determined by the geometry
of spacetime. The renormalization requirements of Wald lead to a an energy
momentum tensor with 2 free parameters which can be conveniently represented
as functional derivatives with respect to the metric of the two quadratic
invariants which one can form from the Ricci tensor and its trace. In
\cite{DFP} the resulting background equations were analyzed in the simpler
conformal limit and it was found that the quantum backreaction stabilizes
solutions i.e. accomplishes a task which usually is ascribed to the
phenomenological cosmological constant. Without the simplifying assumption the
linear dependence on a free renormalization parameter guaranties that any
measured value can be fitted to this backreaction computation. The principles
of QFT cannot determine renormalization parameters.

Hence from a QFT point of view there is no cosmological problem which places
QFT in contradiction with astrophysical observations. A consistency check
would only be possible if there are other measurable astrophysical quantities
which fall into the setting of quantum backreaction on spatially flat RW cosmologies.

\bigskip

\section{Resum\'{e}, additional comments and outlook}

The backbone of this essay has been the quantum counterpart of theories with a
maximal velocity\footnote{Note that in theories with unbounded velocities,
special situations may lead to effective maximal velocities (e.g. velocity of
sound, relativistic particle propagation in DPI).} as compared to those
without. This leads to very different QTs coming with a radically different
localization concept: B-N-W localization related to particles and modular
(causal) localization implemented by local observables. The dividing line, as
the existence of the macrocausal DPI setting shows, is not special relativity
per se but more specifically the (micro)causal nature of the interaction
following from the existence of a maximal velocity.

Both DPI and QFT can be formulated with a joint starting point namely the
one-particle representation spaces as classified by Wigner and their
multiparticle tensor products. Whereas DPI introduces interactions by
modifying the noninteracting n-particle representation "by hand" in such a way
that the modifications of the Poincar\'{e} generators for different n are tied
to each other by cluster factorization, the path from Wigner to QFT is only
simple in the absence of interactions when there is a functorial relation
between modular localized subspaces of the one-particle Wigner space and local
operator subalgebras in a bosonic or fermionic Wigner Fock space. Free fields
display themselves as pointlike coordinatizations of these algebras i.e. as
singular (operator-valued distributions) generators of the net of spacetime
indexed algebras. Whereas the localized one-particle states and the system of
local algebras are unique, there is a countable infinite plurality of relative
local pointlike field coordinatizations which can be divided into two groups,
generators which are linear in Wigner particle creation and annihilation
operators and composites thereof i.e. the Wick ordered monomials of the linear
generators. One can also use stringlike generators (and in some cases there
are no pointlike fields) of which there exist continuously many. The
traditional way to introduce interactions in accord with the locality
principle is by coupling these generating fields "by hand" which is supported
by the non-intrinsic Lagrangian quantization formalism. A truely intrinsic
nonperturbative approach based on the classification of generators for wedge
algebras exists presently only for a subfamily of two-dimensional factorizing models.

Whereas the only localization in the DPI setting is that of B-N-W localized
wave functions in terms of a position operator and its spectral projectors,
the modular localization in LQP leads to dense subspaces and causally complete
subalgebras. Dense subspaces which change with the spacetime region do not fit
into the standard setting of QT in which single operators and projectors on
subspaces play the prominent role. Hence it does not come as a complete
surprise that the literature on this particle-field issue is unfortunately
also somewhat confusing. For many decades QFT was viewed as being part of the
same conceptual setting as QM and only more recently a perception of their
substancial differences developed.

Often deep antagonisms were construed which are really not there. Particles
and fields are in a very precise way asymptotically connected and the B-N-W
localization leads to covariant scattering probabilities for particles
precisely where it is needed, namely for the asymptotic
relation\footnote{Whereas the relation between B-N-W localized events are only
\textit{effectively} covariant at distances larger than a Compton wave length
of the to be localized particles, events seperated by an infinite timelike
distance lead to invariant transition probabilities.} and whether the
coincidence and anticoincidence counters measure asymptotic B-N-W localization
or modular localization is somewhat academic. Instead of dwelling on the lack
of covariance of the B-N-W particle localization it is more realistic to take
the proverbial point of view of the half full glass and emphasize its
\textit{asymptotic} covariance. Without the asymptotic particle concept there
would be no \textit{stability} and \textit{objectivity} since fields are, like
coordinates in geometry, and hence one would not know for sure which one is
being measured. Unlike in classical field theory, where fields have
"individuality" (e.g. electromagnetism), the main role of quantum fields is to
"interpolate" particles and at the some time to implement principles which
cannot be formulated in terms of particles and their S-matrix. Although there
exists an infinite equivalence class of interpolating fields, they interpolate
the same situation; if one wants to avoid the plurality of fields altogether
one can also interpolate the unique system of particles directly with the
unique system of local algebras \cite{Araki}.

Part of the particle-field muddle comes from placing Lagrangian quantization
into the center of QFT and extracting conceptual messages from the proximity
of the Lagrangian quantization formalisms of QM and QFT. This puts a strain on
separating genuinely intrinsic physical properties from those which are
contingent on a particular computational scheme; and it often takes a lot of
thinking to arrive at the conclusion, that besides spacetime-indexed local
algebras on the local level and asymptotic particle states and their
scattering probabilities, there are no intrinsic concepts in QFT; at the end
of the day the classification and construction of models of QFT have to be
understood in terms of these principles and the Lagrangian formalism is only a
temporary crutch.

Without the asymptotic probabilities which only enter through B-N-W
localization, particle physics would not be what it is. QFT in CST in generic
spacetimes (without timelike Killing vectors) lacks these particle concept; as
a consequence of absence of the necessary spacetime symmetries there is no
distinguished vacuum reference state. The question of what remains of particle
physics if a model of QFT admits no standard particle aspects is a very
serious one even in the context of Minkowski space QFT \cite{infra}.

In generic curved spacetime the prerequisite of Poincar\'{e} group
representation theory is absent so that in additions to particles even the
vacuum state has disappeared. This raises the question of what, after
measurements of particles and their scattering cross sections have
disappeared, what remains to be measured at all, is it fields? Certainly there
are still some radiation densities which can be measured in form of the
Hawking/Unruh radiation or cosmic background radiation, but the rich
scattering theory, i.e. particle physics as we have known it for almost 8
decades, does not seem to have a CST counterpart. Wald \cite{recent} has
argued that one should think in terms of measuring fields, but it is not clear
what this means since fields are of a fleeting nature and their stable
non-fleeting aspect consists precisely in their particle/infraparticle
content. In any case the question of what becomes of particles in QFT in CST
is an important open problem.

A large part of this essay was used to expose the existence of two different
kind of \textit{entanglements} which results from a spatial bipartite division
of the global algebra. The standard entanglement picture of quantum
information theory applies only to a quantum mechanical B-N-W localized
subalgebra and its commutant which is B-N-W localized in the complement
region; the result is a tensor factorization and the ensuing notion of
entanglement is that one studied within (quantum) information theory.

In causal QFT such a bipartite division automatically involves the causal
completion of the spatial compact region and its causal disjoint which is the
causal completion of the spatial complement. In this case there is no tensor
factorization and any attempt to go ahead as if it existed leads to divergent
expression; the integrals can be made finite by the use of the practitioner's
sledge hammer: a momentum space cutoff\footnote{Whereas the divergencies in
perturbation theories are not intrinsic i.e. can be avoided by a more
appropriate formulation, the divergencies due to vacuum polarization at the
locus of causal/event horizons are genuine properties of causal QFT which
cannot be "renormalized away".}; but as usual this does not provide any
insight about what is really going on. With some hindsight one may conclude
from such carefully executed cutoff calculations \cite{BKLS} the leading term
of an vacuum polarization-caused area law, but the full insight only results
from answering the question why the tensor factorization fails in the first place.

As well-known the restriction of globally pure state (vacuum, particle states)
to causally localized subalgebras $\mathcal{A(O})$ leads to thermal KMS states
associated with the modular Hamiltonian associated to ($\mathcal{A(O}),\Omega
$). Modular Hamiltonians give rarely rise to geometric movements
(diffeomorphisms). Although in Minkowski spacetime there is no compact
localization region which leads to a \textit{geometric} modular theory, such
situations do occur in connection with appropriate Killing symmetries in CST
if one restricts suitable global states to a black hole
region\footnote{Example: the Hartle-Hawking state on the Kruskal extension
restricted to the region outside a black hole.}.

Although the terminology "entanglement" strictly speaking does not apply to a
bipartite separation with sharp causal boundaries, the literature on
entanglement unfortunately does not differentiate between the QM and the QFT
case (mostly unknowingly, but sometimes also knowingly \cite{Keyl-M}).

We have seen that the split property permits a tensor factorization in which
one tensor factor contains the causally localized algebra, and the second
tensor factor contains its causal disjoint after splitting it away from
spatially touching the original algebra. There is an external parameter
entering the factorization based on the splitting method, namely the splitting
distance $\Delta R.$ The restriction of the global vacuum to one of the tensor
factors is a Gibbs state at a temperature which depends on the normalization
of the modular Hamiltonian which is uniquely associated with this situation.
The entropy of this Gibbs state diverges with decreasing split $\Delta
R\rightarrow0$ and this explains the divergence of the momentum space cutoff
and shows that, different from other divergences in QFT, the localization
entropy in the limit of sharp localization is not a result of a pathology of
the theory or in its formulation but rather the hallmark of a causal QFT with
or without CST. Note that although the split property paves the way for a
return of the entanglement concepts, the homecoming is not complete since the
localization-entanglement is thermal and not information theoretical; there is
simply no way in which a QM bipartite situation can have a thermal
entanglement without having been globally thermal from the start.

The area law and the divergence in the zero split limit hold for localization
with causal horizons as well as for localization behind curved
spacetime-related event-horizons as they occur in black hole physics. This
requires a revision of what is thought to be the interface between QFT in CST
and QG. In many articles the Bekenstein area law (after its quantum
reinterpretation as a property of entropy) was hailed as being part of this
interface since it was not recognized that the \textit{area behavior is a
totally generic behavior of local quantum physics}. To claim that if not the
area law per se than at least Bekenstein's specific gravitational dependent
value is a property of the still illusive QG has a certain plausibility
although it would be the fist time in the history of QT that a classical
constant does not require a quantum modification.

The Bekenstein thermodynamical interpretation of a certain quantity in the
setting of classical gravity raises the question whether it is not possible to
invert this connection i.e. to supplement the thermodynamical setting by
reasonable assumptions of a general geometric nature so that the Einstein
Hilbert equations are a consequence of the fundamental laws of thermodynamics.
Modular theory already relates thermal behavior with localization, hence a
relation of fundamental laws of thermodynamics with gravity is not as
unexpected as it looks at first sight. The reader is referred to some very
interesting observations by Jacobson \cite{Jac}.

Another property attributed to QG is that the event horizon stores a complete
image of the bulk world, i.e. the holography is really an isomorphism. There
are cases in QFT where holography onto a boundary becomes an isomorphism (viz.
the AdS-CFT correspondence) but certainly not on horizons which are
null-surfaces. In the latter case the degrees of freedom on the horizon are
always of a lesser cardinality than those in the bulk and only by enlarging
them by spacetime transformed degrees of freedom outside the null surface can
one return to the bulk. The idea that of a holographic image of the world may
in a future QG setting turning into an isomorphism enjoys some popularity does
not sound very palatable, but it is difficult to criticize something for which
no arguments are given.

As we have seen there is a sharp dichotomy between quantum mechanical
\textit{information theoretical entanglement} and the \textit{thermal
entanglement} resulting from modular localization. Hence it is unclear what
the \textit{black hole information loss} means in the setting of a thermal
localization. In many of the articles the terminology QM instead of QFT is
used, thus making it obvious that the authors do not appreciate the
fundamental differences in the notion of entanglement between QM and QFT.

One of the great advances in reconciling QFT and general relativity is the
discovery of the quantum counterpart of local covariance whose implementation
requires to spacetime-organize an abstract algebraic substrate (e.g. CCR-
CAR-algebra) simultaneously on all possible globally hyperbolic manifolds
together\footnote{Each causally complete submanifold is also an admissible
manifold.} (including of course the Minkowski spacetime) so that the algebraic
substrate on isometrically related manifolds is isomorphic. Since isomorphic
situations cannot be distinguished by experiments within their localization
region, the local covariance principle accomplishes the realization of an
important aspect of \textit{background independence}.

The partisans of quantum gravity think that such situations should not only be
isometric but even identical. In the previous section we have seen that the
local covariance setting has led to the first calculations involving
backreactions in cosmological situations. This makes it possible to address
the problem of the cosmological terms as having its origin in "vacuum" energy
where vacuum in this contest is the euphemistic name for an unknown cosmic
reference state.

Perhaps the most profound difference between QM and QFT finds its expression
in the encoding of a finite number of monads into a certain "modular" position
within a joint Hilbert space. Concretely one thinks of a finite collection of
wedge algebras in certain geometric positions which correspond to nontrivial
modular inclusions and intersections. But the modular positioning is intrinsic
and abstract and in particular does not directly refer to spacetime and its
Poincar\'{e} invariance group. Rather the latter together with a
spacetime-indexed local net of operator algebras is derived from a special
kind of modular positioning. Even the unique nature of the operator algebras
of being monads i.e. hyperfinite type III$_{1}$ algebras is a consequence
since only field theoretic monads allow this positioning. As mentioned in the
section on modular positioning there have been other ideas to highlight the
relational nature of QT in particular Mermin's view of QM in terms of its
correlations as expressed by his apodiction:

\textit{Correlations have physical reality, that what they correlate does not}.

We may express the relational nature of LOP as resulting from modular
positioning as:

\textit{Relative modular positions in Hilbert space have physical reality, the
quantum matter they position does not.}

The presentation of QFT in terms of positioning monads is very specific of LOP
i.e. it has no analog in QM i.e. Mermin's view is not a special case of
positioning in LQP.

Philosophically distinctive viewpoints are however not always the most
appropriate ones for actual constructions. Indeed knowing the action of the
Poincar\'{e} group on one monad (interpreted as a wedge algebra) instead of
the modular positioning of several is the more practical starting point. The
most efficient way to characterize a wedge subalgebra $\mathcal{A}(W)\subset
B(H)$ is in terms of generators. As explained in the paper, in factorizing
two-dimensional models simple generators are known, under suitable conditions
they are Fourier transforms of Zamolodchikov-Faddeev creation/annihilation
operators. In those cases the algebraic construction leads to nontrivial
double cone algebras and finally to the first existence proof of models which
have worse short distance behavior than that allowed by canonical commutation relation.

One would hope for more along these algebraic lines but in view of the fact
that this is the first existence proof in the almost 80 years history of QFT
it should be very encouraging and provide a strong motivation for continuing
along these lines.

There have been similar proposals along modular lines, the most prominent one
being the condition of geometric modular action (CGMA) \cite{CGMA} in which
the modular conjugations rather than the modular groups play the important
generating role. Although they have not been used for the constructions of
models, they proved very useful in the clarification of structural properties
notably the relation of spacetime symmetries with respect to inner unbroken or
spontaneously broken symmetries.

Being interested in the interface between QFT in CST and the still illusive QG
it is natural to ask whether the characterization of QFT in terms of modular
positioning of monads extends beyond Minkowski spacetime. For the simplest
nontrivial kind of QFT, namely chiral theories on a circle, it is well-known
that the Moebius symmetry follows from the modular positioning of two monads,
but it does not lead to more general diffeomorphisms of which none leaves the
Moebius-invariant vacuum fixed. The second message comes from the
representation theory of the Virasoro algebra and states that there can be no
vector at all which is left invariant under a higher diffeomorphism.

Consider for example a diffeomorphism with 4 equally distributed fixed points.
Its geometric aspect leads one to expect a relation with the modular theory of
a 2-interval. It turns out that the modular group of such algebras act in the
expected way as the diffeomorphism with the 4 fix points, but it does do only
on the two interval algebra and not on its complement where its action remains
"fuzzy" i.e. not describable in terms of diffeomorphism. So the message is
that if one admits appropriate vectors which change with the geometrical
situation ("adjusted vacua") and studies the modular theory one can build up
higher diffeomorphism on suitable multi-intervals. The fact that the
diffeomorphism \textit{coalesces with a modular group only on a
multi-interval} is no hindrance.

In view of the nature of local covariance principle such "partial" local
diffeomorphisms together with "partial vacua" seem to be a natural local
generalization of the global vacuum and its associated global symmetries.
Without pressing ahead with the modular positioning approach and reach its
limits, one probably has no chance to get to the interface between QFT in CST
and QG. Whenever one thought to have the first glimpse at QG, as in the
example of the entropic area law or the principle of independence on the
background, one found something in the already existing QFT in CST which put
this view into question. In the case of entropy it was the general area
proportionality, and in the case of background independence the isomorphism
between causally closed parts of different worlds which are diffeomorphic as
manifolds\footnote{It is not clear whether the stronger form of background
independence, in which the isomorphism is replaced by an identity, can be
achieved.}. Thus whenever one deemed to finally have localized the interface
between QFT in CST and QG it volatilized again.

On the other hand there is hardly any doubt that the QM-QFT interface had
reached its conceptual final position. Apart from cosmetic changes one does
not expect major conceptual relocation, even if there remains still a lot of
refurbishing for the quantum measurement and philosophy of science communities.

After completing this essay I became aware of the existence of two papers by
Steve Summers \cite{Sum1}\cite{Sum2} where among other things different
consequences of the split property concerning the localization of spacetime
and inner symmetries are presented. Both papers are a rich source for
additional references.

\end{document}